\documentclass[a4paper,11pt]{article}
\usepackage{jheppub}

\usepackage{graphicx,placeins}
\usepackage{wrapfig}
\usepackage[utf8]{inputenc}
\usepackage{amsmath}
\usepackage{amssymb}
\usepackage{amsfonts}

\newcommand\beq{ \begin{eqnarray} }
\newcommand\eeq{ \end{eqnarray} }

\usepackage{color}

\title{\boldmath Sparse modeling approach to obtaining the shear viscosity from smeared correlation functions }

\affiliation[a]{Department of Physics and Research and Education Center for Natural Sciences, Keio University, 4-1-1 Hiyoshi, Yokohama, Kanagawa 223-8521, Japan}
\affiliation[b]{Department of Mathematics and Physics, Kochi University, 2-5-1 Akebono-cho, Kochi 780-8520, Japan}
\affiliation[c]{Research Center for Nuclear Physics (RCNP), Osaka University, 10-1 Mihogaoka, Ibaraki, Osaka 567-0047, Japan}
\affiliation[d]{
CCSE, Japan  Atomic Energy Agency, 178-4-4, Wakashiba, Kashiwa, Chiba, 277-0871, Japan}
\affiliation[e]{
Mathematical Science Team, RIKEN Center for Advanced Intelligence Project (AIP), 1-4-1 Nihonbashi, Chuo-ku, Tokyo 103-0027, Japan
}

\author[a,b,c]{Etsuko Itou} 
\author[d,e]{and Yuki Nagai}

\emailAdd{itou@yukawa.kyoto-u.ac.jp}
\emailAdd{nagai.yuki@jaea.go.jp}

\abstract{%
We propose the sparse modeling method to estimate the spectral function from the smeared correlation functions.
We give a description of how to obtain the shear viscosity from the correlation function of the renormalized energy-momentum tensor (EMT) measured by the gradient flow method ($C(t,\tau)$) for the quenched QCD at  finite temperature.
The measurement of the renormalized EMT in the gradient flow method reduces a statistical uncertainty thanks to its property of the smearing.
However, the smearing breaks the sum rule of the spectral function and the over-smeared data in the correlation function may have to be eliminated from the analyzing process of physical observables.
In this work, we demonstrate that the sparse modeling analysis in the intermediate-representation basis (IR basis), which connects between the Matsubara frequency data and real frequency data.
It works well even using  very limited data of $C(t,\tau)$ only in the fiducial window of the gradient flow.
We utilize the ADMM algorithm which is useful to solve the LASSO problem under some constraints.
We show that the obtained spectral function reproduces the input smeared correlation function at finite flow-time.
Several systematic and statistical errors and the flow-time dependence are also discussed.
}

\begin{document} 
\maketitle
\flushbottom

\section{Introduction}
\label{sec:intro}
One of the most important results obtained by RHIC (Relativistic Heavy Ion Collider) experiment is an elliptic flow of QCD particles~\cite{Adams:2005dq,Adcox:2004mh}.
The elliptic flow data using a Boltzmann-type equation for gluon
scattering indicates a large cross section above the phase transition
temperature ($T_c$)~\cite{Molnar:2001ux}.
Furthermore, this elliptic flow is well explained by the hydrodynamics of the quark-gluon-plasma (QGP) state~\cite{Teaney:2000cw,Kolb:2000fha,Huovinen:2001cy,Teaney:2001av,Hirano:2001eu,Hirano:2002ds,Teaney:2003kp,Teaney:2009qa}.
Thus, the QGP state is not described by a free gas system of perturbed gluons.
It might be natural since around $T_c$ the interaction force between the quarks and gluons becomes strong.
On the other hand, the numerical simulation of the relativistic liquid system gives that the upper limit of the ratio between the shear viscosity ($\eta$) to the thermal entropy ($s$)  is very small, $\eta/s  \le 0.4$~\cite{Teaney:2009qa}.
It turns out that the QGP has the perfect liquid property. 
The result gives us a non-sticky image of QGP states even though the QCD particles are strongly interacting.
To make these curious properties of QGP clear,  a considerable number of studies have been conducted on the determination of the shear viscosity from the first principle calculation in the pure SU($3$) gauge theory~\cite{ Aarts:2002cc, Nakamura:2004sy,Meyer:2007ic,Moore:2008ws, Meyer:2008dq, Meyer:2009jp, Meyer:2011gj, Mages:2015rea,Astrakhantsev:2017nrs, Astrakhantsev:2018oue,Pasztor:2018yae}.

The shear viscosity is given by the first-differential coefficient of the spectral function at zero frequency ($\eta (T) = \pi d \rho (\omega)/ d \omega $ at $\omega=0$).
The spectral function is defined by the Euclidean correlation function of the renormalized spacial energy-momentum tensor (EMT) in the static state as follows:
\beq
C(\tau) &=& \frac{1}{T^5} \int d \vec{x} \langle T^R_{12} (0, \vec{0}) T^R_{12}(\tau, \vec{x}) \rangle= \int^{+\infty}_{-\infty} d \omega K(\tau, \omega) \rho(\omega).\label{eq:fisrt-eq}
\eeq 
Here, $K(\tau,\omega)$ denotes the kernel of an integral transform.
In the lattice simulations, $C(\tau)$ is measured by generated configurations using the first equality sign, and then $\rho(\omega)$ is estimated through the second one.

During these calculations, there are at least three difficulties to obtain $\rho(\omega)$:
(i) How to define the renormalized EMT on the lattice (ii) how to improve a bad signal-to-noise ratio of the correlation function of EMT (iii) how to estimate $\rho(\omega)$ from the limited number of the data $C(\tau)$ on the lattice.

As for the first and second problems, a promising method, called the gradient flow method, has been provided in Ref.~\cite{Suzuki:2013gza,Asakawa:2013laa}.
By using the UV finiteness of the flowed operators in the gradient flow~\cite{Luscher:2010iy,Luscher:2011bx},
we can obtain the renormalized EMT without hard numerical calculation of $Z$-factor for quenched QCD.
Furthermore, thank  the smearing property of the gradient flow, the statistical uncertainty is suppressed in this method.
On the other hand, it is notable that we have to analyze the data only in the fiducial window of the flow-time to obtain the thermodynamic quantities using the gradient flow method in Ref.~\cite{Asakawa:2013laa}.
The range is given by $2a \ll \sqrt{8t} \ll T^{-1} = N_\tau a$, where both a strong lattice discretization and  over-smearing corrections are suppressed.

The third problem is well known as an ill-posed inverse problem. 
Especially in the finite temperature system, the number of sites in the temporal direction is very limited, then it makes harder to solve the inverse problem.
\begin{figure}[h]
\begin{center}
\includegraphics[scale=0.45]{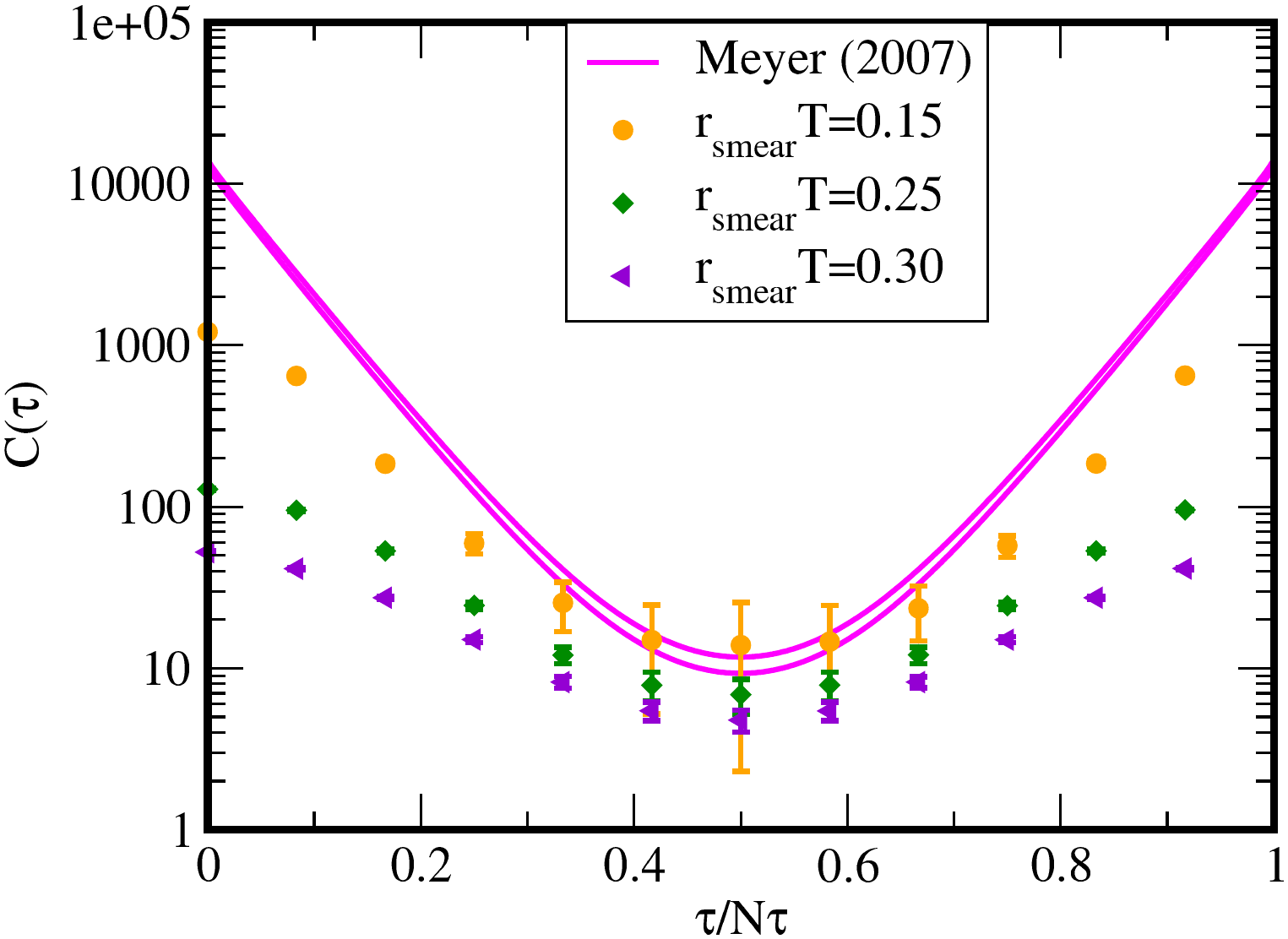}
\caption{Flow-time dependence of the $T_{12}$ correlation function. Here, the data of $\beta=6.72, N_s^3\times N_\tau=48^3 \times12$ ($T=1.65T_c$) are plotted. Magenta curves denotes the best-fit function with 1-$\sigma$ error of the fitting parameters which is reconstructed by the estimated spectral function by using the Breit-Wigner ansatz given in Ref.~\cite{Meyer:2007ic} . It indicates a non-smeared correlation function. $r_{\mbox{smear}}T=\sqrt{8t}/aN_\tau$, where $t$ denotes the flow-time. This plot is firstly shown in Ref.~\cite{Itou:2017ypy}.
}
\label{fig:T12T12-corr-flowtime-deps}
\end{center}
\end{figure}
If we use the gradient flow method to obtain the correlation function, taking the data only in the fiducial window makes the third problem worse, since the available number of the data is further limited.
To see the origin of the demerit, we show the smeared correlation function $C(t,\tau)$ in Fig.~\ref{fig:T12T12-corr-flowtime-deps}. 
We find that 
the slope of $C(t,\tau)$ in the short (and long because of the periodicity) $\tau$-regime shows the discrepancy from the non-smeared data presented in Ref.~\cite{Meyer:2007ic}.
The discrepancy becomes large if the distance of two operators ($\tau$) gets shorter than the smeared length ($\sqrt{8t}$) by the gradient flow since the smeared regime of one operator reaches at the position of the other operator in the correlation function.

Now, a question arises: whether we can straightforwardly apply ordinary estimation methods of $\rho(\omega)$ to such a smeared correlation functions. 
Until now, the fitting with some ansatz ({\it e.g.}, the Breit-Wigner ansatz) and the Bayesian methods ({\it e.g.}, the maximal entropy method~\cite{Nakahara:1999vy,Asakawa:2000tr}) have been utilized for the  estimation of the spectral function in the case of the non-smeared correlation function, $C(\tau)$.
 For the smeared EMT correlation function obtained by the gradient flow method in $N_f=2+1$ QCD, the fitting analyses with the Breit-Wigner ansatz and the hard thermal loop ansatz for $\rho(\omega)$ have been applied~\cite{Taniguchi:2019eid}.
The authors have tried to fit the data of $C(t,\tau)$ only in the limited $\tau$-regime to eliminate the over-smeared data. 
However, the analysis for the eliminated $C(t, \tau)$ using the same functional form of $\rho(\omega)$ is questionable since originally the non-smeared $C(\tau)$ and $\rho(\omega)$ are related via the integration equation Eq.(\ref{eq:fisrt-eq}) and $C(\tau)$ at each $\tau$ gets the corrections from all $\omega$-regime.
Thus, the lack of $C(t,\tau)$ at several $\tau$ may change the functional form of the spectral function.
As the other directions, several methods to reconstruct the spectral function with smearing have been also proposed in Refs.~\cite{Burnier:2013nla,Hansen:2017mnd,Tripolt:2018xeo,Hansen:2019idp,Kades:2019wtd, Bailas:2020qmv}.

In this work, we propose the sparse modeling method to estimate the spectral function at finite flow-time, which is based on Bayesian statistics.
Here, we need not introduce any functional form of the spectral function.
The sparse modeling method can be applied to both non-smeared and smeared correlation functions.
It has been utilized in board topics, {\it e.g.} MRI~\cite{MRI} and taking a picture of a large blackhole~\cite{Blackhole} (see also a recent review~Ref.~\cite{review-sparse}).
As for the estimation of spectral function, it has been succeeded for several analyses in the many-body system~\cite{Shinaoka2017a, Shinaoka2017b}.
Utilizing the intermediate-representation basis (IR basis)~\cite{Shinaoka2017a,Nagai2018,Chikano2018a,Chikano2018b,Li2019} and a few reasonable constraints, the sparse modeling makes it possible to estimate the real-frequency representation of $\rho(\omega)$ from a sparse imaginary-time correlation function.

We formulate the sparse modeling analysis for $C(t,\tau)$ only in the fiducial window.  
Here, we carry out  the singular value decomposition (SVD) of the kernel $K(\tau,\omega)$ in the IR basis, which is independent of the Monte Carlo data.
The singular values of the kernel decay exponentially or even faster, then we drop the degree of freedom for the correlation function and the spectral function in the IR basis associated with the sufficiently small singular values.
Then, we find the spectral function to minimize the square error given by the difference between the input correlation function and the output one reconstructed by $\rho(\omega)$.
During this process, the $L_1$ regularization term is added to be consistent with the truncation of the degree of freedom in the IR basis.
Technically, we utilize the ADMM algorithm to solve the optimization problem with such a regularization term, which is recently proposed by  S.~Boyd {\it et al.}~\cite{ADMM1, ADMM2}.
In this work, we demonstrate that the sparse modeling analyses using very limited data  for the quenched QCD at  finite temperature.

This paper is organized as follows: 
In \S.~\ref{sec:sparse-modeling}, we give a review of the sparse modeling method. 
In \S.~\ref{sec:EMT}, we explain how to calculate the correlation function of the renormalized EMT for the quenched QCD in the gradient flow method.
 In \S.~\ref{sec:sparse-Ctau}, we give a standard description to obtain the shear viscosity, where we assume that the correlation function in the whole $\tau$-regime is available for the sparse modeling analysis.  
 Then, we modify the standard description to analyze the smeared correlation function in which the over-smeared data are eliminated in \S.~\ref{sec:tech-step}.
Section~\ref{sec:setup} presents the simulation setup of this work.
We show the result of the sparse modeling analysis using the central value of the smeared correlation function in \S.~\ref{sec:SpM}.
We find that it works well since the reconstructed correlation function from the obtained spectral function is almost consistent with the input data.
\S.~\ref{sec:error} contains the error estimations.
We investigate several systematic errors; $\omega_{cut}$, $\tau$-regime, and the flow-time dependence in our analysis, and also estimate the statistical uncertainty..
Although the statistical error of our result is sizable, we show that the bootstrap analysis is promising for estimating the statistical error of this analysis, since an expected relationship between the correlation function and the spectral function in the IR basis is satisfied.
In this work, we have only $2,000$ configurations for the measurement of the correlation function, and it is very few in comparison with $800,000$ and $6$ million configurations in Refs.~\cite{Nakamura:2004sy,Pasztor:2018yae}.
Judging from such a poor statistic, we will give up the precise determination of the shear viscosity and focus on the feasibility of the sparse modeling method to obtain the spectral function combing with the gradient flow method.
The last section is devoted to the summary.

\section{Sparse modeling method}\label{sec:sparse-modeling}
Here, we give a brief review of the sparse modeling method in the IR basis following Refs.~\cite{Shinaoka2017a, Shinaoka2017b} (see also a review paper~\cite{review-sparse}). 

The input is the imaginary-time Green's function $C(\tau)$.
Its Fourier transform $C(i \omega_n)$, where $\omega_n$ denotes a Matsubara frequency, is related with 
the spectral function $\rho(\omega)$ as $\rho(\omega)= \frac{1}{\pi} \mbox{Im} C(\omega + i 0)$ by replacing $i \omega_n$ with $\omega + i0$.
Then, the input $C(\tau)$ is written by the integral form of the spectral function,
\beq
C(\tau) = \int^{+\infty}_{-\infty} d \omega K(\tau, \omega) \rho(\omega), \label{eq:def-C}
\eeq
where $0 \le \tau \le 1/T$.
The kernel $K$ in this work is given by
\beq
K(\tau, \omega) = \frac{\cosh \left( \omega (\frac{1}{2T}-\tau) \right)}{\sinh (\frac{\omega}{2T})},
\eeq
but in this section the discussion does not depend on the details of the kernel.
Here, we assume that the kernel is described by some exponential function as in many systems.

For simplicity,  we denote Eq.(\ref{eq:def-C}) using the dimensionless vectors as
\beq
\vec{C} \equiv K \vec{\rho}. \label{eq:ckrho}
\eeq
The component of the vector $\vec{C}$ is $C_i \equiv C(\tau_i)$, where $\tau_i$ labels the temporal site on the lattice with $0 \le \tau_i \le N_\tau-1$.
The kernel $K$ becomes $N_\tau \times N_\omega$ matrix, while $\vec{\rho}$ denotes a vector whose component is $\rho_j \equiv \rho(\omega_j)$ with $j = 1, \cdots ,N_\omega$.

Our goal is to find $\vec{\rho}$ to minimize the square error
\beq
\chi^2 (\vec{\rho}) = \frac{1}{2} \| \vec{C} - K \vec{\rho}  \|^2_2.\label{eq:chi-sq}
\eeq
Here, $\| \cdot \|_2$ stands for the $L_2$ norm defined by $\| \vec{\rho} \|_2 \equiv (\sum_j \rho^2_j)^{1/2}$.
Note that the vector $\vec{C}$ is sparse, so that the amount of the information in $\vec{C}$ is much smaller than that in $\vec{\rho}$.
The fact leads to the instability of $\vec{\rho}$.

We have only the Monte Carlo data of $C(\tau)$, which is not the same with the true value of $\vec{C}$ and the data have the statistical uncertainty.
Here, we call the discrepancy between the Monte Carlo data and the true value of $\vec{C}$ a ``noise".

Equation~(\ref{eq:ckrho}) gives simultaneous linear equations, where the number of equations is $N_\tau$ while one of the unknown variables is $N_\omega$.
We can solve the equation if $N_\tau = N_\omega$ and the kernel is given by a regular matrix, while the unique solution does not exist and several possible solutions are allowed in the case of $N_\tau < N_\omega$.

Here, we would like to choose a stable solution against the noise among these possible solutions.
For this purpose, we take an efficient basis called an IR basis.
By introducing the singular value decomposition (SVD) of the matrix $K$ defined as 
\beq
K = U S V^\dag.
\eeq
We rewrite Eq.(\ref{eq:ckrho}) as 
\beq
\vec{C} \equiv  U S V^\dag \vec{\rho},
\eeq
where $S$ is an $N_\tau \times N_\omega$ diagonal matrix, and $U$ and $V$ are unitary matrices of size $N_\tau \times N_\tau$ and $N_\omega \times N_\omega$, respectively.
The new vectors in the IR basis
\beq
\vec{\rho}' \equiv V^t \vec{\rho}, ~~~~~~~~ \vec{C}' \equiv U^t \vec{C}.\label{eq:rho-C-IR-basis}
\eeq
imply the square error Eq.(\ref{eq:chi-sq}) as
\beq
\chi^2(\vec{\rho}) = \frac{1}{2} \| \vec{C}' - S \vec{\rho}' \|^2_2 = \frac{1}{2} \sum_l (C'_l - s_l \rho'_l)^2.\label{eq:chi-sq-SVD}
\eeq
Thus, at the minimum point of the square error the $l$-th element  almost satisfies 
\beq
 [\vec{C}']_l =  s_{l} [\vec{\rho}']_l.\label{eq:Cl-sl-rhol}
\eeq
It turns out that the contribution of $\rho'_l$ to $\chi^2 (\vec{\rho}')$ is weighted by the corresponding $s_l$.
The singular values $s_l$ with $l = 1,2,\cdots $ of this type of kernels decay exponentially or even faster.
Then, for example, if double precision numbers are used, $C'_l$ can not include the information about $\rho'_l$, where $s_{l}/s_{1}$ is less than $10^{-16}$ on a computer.
It {\it does not } depend on simulation details since a kernel $K$ is fixed. 
This fact naturally  explains the reason why a tiny noise of $C(\tau)$ leads to a large difference of $\rho(\omega)$. 
Therefore, a tiny fluctuation in the left-hand side can affect the ``best" inference of $\rho(\omega)$, and several ``likely" solutions satisfy Eq.~(\ref{eq:def-C}).

In actual calculations, the component  $s_l\rho'_l$ with sufficiently small singular values  gives a negligible contribution to $\chi^2 (\vec{\rho})$.
Then, by introducing some threshold $s_{cut}$, we can drop such a component $l > l_{cut}$, where $s_{l} < s_{cut}$. 
This truncation leads us to obtain a stable solution which is robust against the noise of $C(\tau)$.
\begin{figure}[h]
\begin{center}
\includegraphics[scale=0.4]{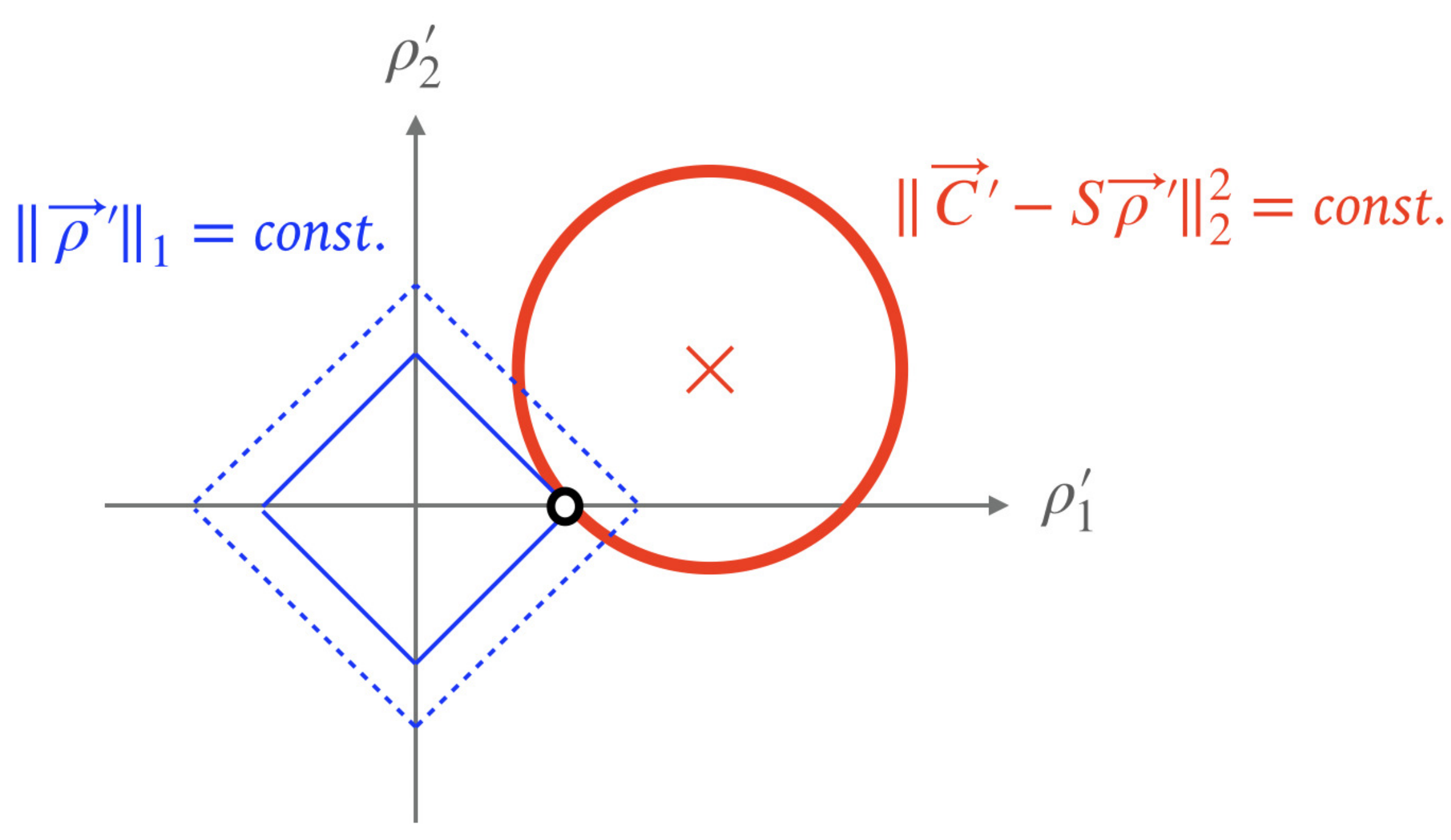}
\caption{Image of $L_1$ and $L_2$ regularizations in 2-dimensional $\rho'_l$ plane. The $L_1$ ($L_2$) regularization can be described by a line (a circle).
The intersection point (open symbol) appears on the horizontal axis, where $\rho'_2$ is zero.
}
\label{fig:L1-reg}
\end{center}
\end{figure}

Now, the vectors in the IR basis are reduced to the $l_{cut}$-component vectors.
Therefore,  we would like to find the solution $\rho'_l$  in Eq.~(\ref{eq:chi-sq-SVD}), where  many components of $\vec{\rho}'$ takes zero to be consistent with the truncation.
To search for such a solution $\rho'_l$, we consider the cost function with an $L_1$ regularization term,
\beq
F(\vec{\rho}') \equiv \frac{1}{2} \| \vec{C}' - S \vec{\rho}'  \|_2^2 + \lambda \| \vec{\rho}' \|_1,\label{eq:cost-fn-L1}
\eeq 
where $\lambda$ is a positive constant and $\| \cdot \|_1$ denotes the $L_1$ norm defined by
\beq
\| \vec{\rho}' \|_1 \equiv \sum_l | \rho_l'|.
\eeq
This $\lambda$ plays the role of the Lagrange multiplier.
If the value of $\lambda$ is smaller than the contribution of the noise for $\vec{C}$, then the obtained $\vec{\rho}$ must be consistent with the true spectral function within the uncertainty coming from the noise.

An intuitive picture of the role of the $L_1$ regularization is following.
The $L_1$ regularization ($\| \vec{\rho}' \|_1 = $const.) gives a linear constraint for the components of $\vec{\rho}'$, and is graphically drawn by dotted and solid lines in Fig.~\ref{fig:L1-reg}. 
The minimize problem of the cost function turns  to tune $\lambda$ to minimize the sum of two constants: $\|  \vec{C}' - S \vec{\rho}'    \|_2^2 $ is a constant and $\| \vec{\rho}' \|_1  $ is the other constant.
In Fig.~\ref{fig:L1-reg}, we see that the open-circle point becomes the most favorable.
Note that at the open-circle point, $\rho_2$ vanishes and the number of the effective components is reduced.

Actually, according to the example analyses in Ref.~\cite{Shinaoka2017b, review-sparse}, the spectral function becomes featureless for $\lambda > \lambda^{opt}$, while artificial spikes appear for $\lambda < \lambda^{opt}$.
In other words, the former case corresponds to the under-fitting, where the $L_1$ regularization term is too strong and the number of components  $\rho'_l$ is too reduced.
On the other hand, the latter case does to the over-fitting, where the $L_1$ term is too weak and the vector $\vec{\rho}'$ has redundancy.

The optimization problem with these $L_1$ and/or $L_2$ regularization is called  the LASSO (Least Absolute Shrinkage and Selection Operators) problem~\cite{LASSO}.
It allows obtaining the global minimum regardless of initial conditions~\cite{ADMM1, ADMM2}.
As an additional constraint, we will add the non-negativity condition of the spectral function,
\beq
\rho(\omega) \ge 0,
\eeq 
in this work.
Furthermore, we can also use the sum rule,
\beq
\sum_j \rho_j=1,
\eeq
as an additional condition, while this constraint is not  used in this work.
The numerical algorithm to solve the optimization problems with these constraints has been developed~\cite{ADMM1, ADMM2}, and it is called the alternating direction method of multipliers (ADMM) algorithm.
See Appendix~\ref{sec:ADMM} for the detail  of the algorithm.

\section{Shear viscosity in quenched QCD}\label{sec:shear-viscosity}
In this section, we explain how to obtain the shear viscosity for the quenched QCD in the lattice simulation. 
Firstly, we explain the calculation strategy of the correlation function of EMT using the gradient flow.
Secondly, we give a standard formula for the sparse modeling method. 
Here, we assume that the correlation function in the whole $\tau$-regime is available for the sparse modeling analysis.
Then, we modify the standard formula for the smeared correlation function at  finite flow-time, where we eliminate the over-smeared data.

\subsection{Measurement of the correlation function of EMT in the gradient flow method}\label{sec:EMT}
The renormalized EMT for the quenched QCD in the gradient flow method~\cite{Suzuki:2013gza} is given by 
\begin{align}
   T_{\mu\nu}^R(x)
   =\lim_{t\to0}\left\{\frac{1}{\alpha_U(t)}U_{\mu\nu}(t,x)
   +\frac{\delta_{\mu\nu}}{4\alpha_E(t)}
   \left[E(t,x)-\left\langle E(t,x)\right\rangle_0 \right]\right\},
\label{eq:(4)}
\end{align}
at a flow-time $t$.
Here, we utilize the small flow-time expansion and drop ${\cal O} (t)$ corrections.
The coefficients $\alpha_U,\alpha_E$ are calculated perturbatively in Ref.~\cite{Suzuki:2013gza}.
An advantage of the usage of the gradient flow method is the absence of $Z$-factor to define the renormalized EMT.
Once we use the renormalized coupling constant in the coefficient $\alpha_U, \alpha_E$, all composite operators constructed by the flowed gauge field are the UV finite at positive flow-time ($t>0$)~\cite{Luscher:2011bx}.
The last term in right-hand-side, $\langle\cdot\rangle_0$, describes the vacuum expectation value (v.e.v.), which relates to the zero point energy. 
$U_{\mu \nu}(t,x)$ and $E(t,x)$ denote gauge-invariant local products of dimension~$4$.

In the continuum theory,  
$U_{\mu\nu}(t,x)\equiv G_{\mu\rho}(t,x)G_{\nu\rho}(t,x)
-\delta_{\mu\nu} E(t,x)$
and~$E(t,x)\equiv\frac{1}{4}G_{\mu\nu}(t,x)G_{\mu\nu}(t,x)$~\footnote{Here, we drop the color indices.}.
Here, $G_{\mu \nu}$ represents the field strength constructed by the flowed gauge field ($B_{\mu}(t,x)$).
This flowed gauge field is a solution to the gradient flow equation,  
\beq
\partial_t B_{\mu} = D_{\nu} G_{\nu \mu}, ~~~~ B_{\mu} (t=0,x) = A_{\mu}(x),\label{eq:gflow}
\eeq
where $A_{\mu}(x)$ denotes the original quantum gauge field variable.
The Fourier components $B_\mu (p)$ have a suppression factor $e^{-p^2t}$.
 Therefore, the flow-time ($t$) plays a role of the UV cutoff in momentum space. 
In the coordinate space, the flowed gauge field can be interpreted by the smeared field in the range of $|x| < \sqrt{8t}$.

On the lattice, $G_{\mu \nu}$ operator can be calculated by the clover-leaf operator, whose size is $2a$.
The relationship Eq.~(\ref{eq:(4)}) is useful in the range of $2a < \sqrt{8t}$, where the corrections of the lattice discretization  would be mild.

The numerical calculation using this formula has firstly done in Ref.~\cite{Asakawa:2013laa}, where the expectation value of the EMT has been measured for the quenched QCD at finite temperature.
The expectation value of EMT is directly related to the bulk quantities, {\it e.g.} integration measure and thermal entropy.
The results are consistent with the ones obtained by the integration method within the statistical error.
Thanks to the smearing effect of the flowed fields, the statistical uncertainty is smaller than the other method.
On the other hand, we need to take a fiducial window of the flow-time, $2a < \sqrt{8t} < aN_\tau/2$, to avoid a strong lattice discretization and  over-smearing corrections.

Now, we calculate the two-point function of EMT.
It is related to the shear and bulk viscosities.
In this work, we focus on the former one, which is given by the correlation function of $T^R_{12}$ component.
The Euclidean correlation function is defined as
\beq
C(\tau) = \frac{1}{T^5} \int d \vec{x} \langle T^R_{12} (0) T^R_{12}(x) \rangle,\label{eq:def-Ctau-latt}
\eeq 
where $x=(\tau, \vec{x})$.

On the lattice at a finite flow-time $t$, we measure the two-point function of $U_{12} (t,x)$,
\beq
C (t,\tau/a)= N_\tau^5 \sum_{\vec{x}} \frac{1}{\alpha_U(t)^2} \langle U_{12} (t, 0) U_{12}(t,x) \rangle,\label{eq:def-smear-C}
\eeq
where the second argument of $U_{12} (t,x)$ denotes a four-dimensional vector $x=(\tau/a, \vec{x}/a)$.
Although in an ideal double limits, $a \rightarrow 0$ and then $t\rightarrow 0$ limits, this smeared correlation function goes to the Euclidean correlation function (Eq.~(\ref{eq:def-Ctau-latt})), practically it is better to calculate physical observables from the smeared correlation function and take the double limits for the observables directly.
In this work, we would like to obtain the shear viscosity from $C(t,\tau/a)$ at finite flow-time.
How to treat $C(t,\tau/a)$ in the sparse modeling analysis will be discussed in \S.~\ref{sec:tech-step}.

\subsection{Estimation of the spectral function using the sparse modeling method}\label{sec:sparse-Ctau}
Here, we give a standard formula for the sparse modeling method.
We assume that $C(\tau)$ is obtained after the double limits or $C(t,\tau)$ in the whole $\tau$-regime is available for the analysis.
In this section, we simply denote these correlation functions as $C(\tau)$. 

The correlation function can be expressed in terms of the corresponding spectral functions ($\rho(\omega)$) as
\beq
C(\tau) = \frac{1}{T^5} \int_{0}^{\infty} d \omega \rho(\omega) \frac{\cosh \omega \left( \frac{1}{2T} - \tau \right)}{\sinh(\frac{\omega}{2T})}.\label{eq:C-rho-cont}
\eeq
The spectral function has the following properties,
\beq
\frac{\rho(\omega)}{\omega} \ge 0, \quad \rho(-\omega) = -\rho(\omega).
\eeq
The shear viscosity is given by 
\beq
\eta (T) = \pi \frac{d \rho}{d \omega} |_{\omega=0},\label{eq:def-shear}
\eeq
from the spectral function.

On the lattice, Eq.~(\ref{eq:C-rho-cont}) turns to
\beq
C(\hat{\tau})&=& N_\tau^5 \int^{\infty}_0 d \hat{\omega}  \hat{\rho}( \hat{\omega}) \frac{\cosh \hat{\omega} \left( \frac{N_\tau}{2} -\hat{\tau} \right)}{\sinh (\frac{\hat{\omega} N_\tau }{2})},\label{eq:EMT-corr-org}
\eeq
where $\hat{\tau}=\tau/a$, $\hat{\omega}=a \omega$ and $\hat{\rho} = a^4 \rho $, respectively.

Equation (\ref{eq:def-shear}) shows that the shear viscosity is the first-differential coefficient at $\omega=0$, then it is convenient to define
\beq
\tilde{\rho}(\hat{\omega}) \equiv \frac{\hat{\rho} (\hat{\omega})}{2  \tanh (\hat{\omega} N_\tau/2)} \label{eq:rho-tilde}.
\eeq
Note that $\tilde{\rho}(\hat{\omega})$ turns to an even-function in terms of $\hat{\omega}$.
Then, the relationship with the correlation function (\ref{eq:EMT-corr-org}) is expressed by
\beq
C(\hat{\tau}) = N_\tau ^5 \int_{-\infty}^\infty d \hat{\omega} \tilde{\rho}(\hat{\omega}) \frac{\cosh(\hat{\omega} \left( \frac{N_\tau}{2} -\hat{\tau} \right))}{ \cosh ( \hat{\omega} N_\tau/2 )}.\label{eq:C-tilde-rho}
\eeq

To obtain the spectral function in Eq.(\ref{eq:C-tilde-rho}), we introduce the cutoff of $\omega$ and the rescaled $\tau$ as follows:
\beq
\omega' \equiv \frac{\omega}{\omega_{cut}} = \frac{\hat{\omega}}{\hat{\omega}_{cut}}, ~~~ \tau' \equiv \frac{\tau}{aN_\tau}=\frac{\hat{\tau}}{N_\tau}.
\eeq
Here, the regimes of the primed variables are $-1 \le \omega' \le 1$ and $ 0 \le \tau' < 1$.
Using $\Lambda = N_\tau \hat{\omega}_{cut}$, the relationship between the correlation function and the spectral function is rewritten by
\beq
C(\tau')&=&N_\tau^4 \Lambda \int_{-1}^1 d \omega ' \tilde{\rho}(\omega ') \frac{\cosh [\frac{\omega ' \Lambda }{2} (2 \tau' -1)]}{\cosh (\omega ' \Lambda /2)}.
\eeq

Discretizing the integral into a finite sum and taking a replacement $ (d\omega ' ) \rightarrow \Delta \omega'$ with $\Delta \omega ' = 2/(N_{\omega}-1)$,
the finite sum is given by~\footnote{In the quadrature by parts, the upper value of the sum is ($N_\omega-1)$ in Eq.~(\ref{eq:Nomega-sum}), here we take $N_\omega$ in our analysis. 
As a result, the integrand $\rho(\omega)$ is approximately zero at $\omega = \pm \omega_{cut}$, so that practically the contribution of $N_\omega$-th data to the sum is negligible.}
\beq
\frac{C(\tau_i')}{N_\tau ^4 \Lambda} = \sum_{j}^{N_\omega} (\Delta \omega ' ) \tilde{\rho}(\omega_j')  \frac{\cosh [\frac{\omega_j ' \Lambda }{2} (2 \tau_i' -1)]}{\cosh (\omega_j ' \Lambda /2)}.\label{eq:Nomega-sum}
\eeq
To make the $N_\omega$ dependence mild in the sparse modeling analysis, we include $\sqrt{\Delta \omega'}$ in the kernel as follows:
\beq
K(\tau'_i,\omega'_j)  \equiv \frac{\cosh [\frac{\omega_j ' \Lambda }{2} (2 \tau'_i -1)]}{\cosh (\omega_j ' \Lambda /2)} \sqrt{\Delta \omega'}.
\eeq
Note that the kernel at $\omega'=0$ is independent of $\tau$-coordinate.

Now, the equation we have to solve is
\beq
\frac{C(\tau'_i)}{N_\tau^4 \Lambda} &=& \sum_j^{N_\omega} K(\tau'_i,\omega'_j) \tilde{\rho}_{new} (\omega_j ')~~~ \mbox{with}~~~
\tilde{\rho}_{new} (\omega_j ') = \tilde{\rho} (\omega_j ') \sqrt{\Delta \omega'}. \label{eq:C-rho-new}
\eeq
The spectral function $\tilde{\rho}_{new}$ satisfies the sum rule
\beq
\sum_j^{N_\omega} \tilde{\rho}_{new} (\omega_j ') =\frac{C(\tau_0')}{N_\tau^4 \Lambda \sqrt{\Delta \omega '}},
\eeq
which is given by Eq.~(\ref{eq:C-rho-new}) at $i=0$.
Therefore, a normalized spectral function,
\beq
\tilde{\rho}_{calc}  \equiv \tilde{\rho}_{new}  \left[ \frac{C(\tau'_0)}{N_\tau^4 \Lambda \sqrt{\Delta \omega '}} \right]^{-1},
\eeq
implies a simple form of the sum rule,
\beq
\sum_j \tilde{\rho}_{calc} (\omega_j ') = 1.\label{eq:sum-rule-Sp}
\eeq
We will obtain this spectral function $\tilde{\rho}_{calc}$ as an output of the sparse modeling analysis.

The original spectral function $\tilde{\rho}(\omega'_j)$ in Eq.~(\ref{eq:Nomega-sum}) is transformed from the output of the sparse modeling analysis ($\tilde{\rho}_{calc}$),
\beq
\tilde{\rho} (\omega_j ') &= &\frac{\tilde{\rho}_{new}}{\sqrt{\Delta \omega '}} 
= \frac{C(\tau'_0)}{N_\tau^4 \Lambda \Delta \omega '} \tilde{\rho}_{calc} (\omega_j ').\label{eq:tilde-rho-calc}
\eeq
Finally, the shear viscosity divided by $T^3$ to be a dimensionless quantity is given by 
\beq
\frac{\eta}{T^3} &\equiv & \pi \frac{1}{T^3} \frac{d \rho (\omega)}{d \omega} |_{\omega =0} \nonumber\\
&=& \pi {N_\tau^4} \tilde{\rho} (0). \label{eq:eta-rho-calc}
\eeq

Let us summarize the technical steps of our strategy to obtain the shear viscosity using the sparse modeling method as follows:

{\bf Step 1:} Measure the correlation function $C(\tau_i)$ on the lattice, and then normalize it using the value of $C(\tau_i=0)$.
\beq
C_{calc} (\tau_i') &=& \frac{C(\tau'_i)}{N_\tau^4 \Lambda} \left[  \frac{C(\tau'_0)}{N_\tau^4 \Lambda  \sqrt{\Delta \omega'}}   \right]^{-1}= \frac{C(\tau_i')}{C(\tau'_0)} \sqrt{\Delta \omega '}.\label{eq:norm-C}
\eeq

{\bf Step 2:} Construct the discretized kernel
\beq
K(\tau'_i,\omega'_j)  \equiv \frac{\cosh [\frac{\omega_j ' \Lambda }{2} (2 \tau_i' -1)]}{\cosh (\omega_j ' \Lambda /2)} \sqrt{\Delta \omega'}\label{eq:kernel-SVD}
\eeq
and carry out SVD decomposition. Here, we utilize DGESDD routines of LAPACK.

{\bf Step 3:} Estimate an optimal value of $\lambda$ as follows:
We first fix a searching range of $\lambda$ as $[\lambda_{min},\lambda_{max}]$.
Next, we define a line segment function in log scale, $f(\lambda)=a \lambda^b$, which connects $f(\lambda_{mix})$ with $f(\lambda_{max})$.
Then, we search $\lambda^{opt}$, where the ratio $f(\lambda)/\chi^2(\vec{\rho})$ has a peak since it corresponds to the position of kink in $\chi^2 (\vec{\rho})$. 
Furthermore, a simple check of whether an obtained $\lambda$ is a ``correctly" optimized value or not can be done by seeing the scaling property of $\lambda^{opt}$.
The ADMM algorithm has free parameters ($\mu, \mu'$).
The correct $\lambda^{opt}$ is inversely scaled by the values of ($\mu$, $\mu'$) in the ADMM algorithm (see Eq.~(\ref{eq:ADMM-nu}) in Appendix~\ref{sec:ADMM}).

{\bf Step 4:} Carry out the sparse modeling analysis using the ADMM algorithm to find the most likely spectral function $\tilde{\rho}_{calc}$ using $\lambda^{opt}$.
We can  reconstruct the correlation function using the obtained $\tilde{\rho}_{calc}$ as 
\beq
C_{output}(\tau'_i) = \frac{C(\tau'_0)}{\sqrt{\Delta \omega}} \sum_j K(\tau'_i, \omega'_i) \tilde{\rho}_{calc} (\omega'_i) .
\eeq
We check the feasibility of the analysis whether  $C_{output} (\tau_i)$ reproduces the input correlation function $C(\tau_i)$.

{\bf Step 5:} Calculate the shear viscosity ($\eta(T)/T^3$) on the lattice from the obtained spectral function using Eqs.~(\ref{eq:tilde-rho-calc}) and (\ref{eq:eta-rho-calc}).
Carry out the same calculations from {\bf Step 1} to {\bf Step 4} using the other lattice spacings with keeping a temperature, and then obtain $\eta(T)/T^3$ in the continuum extrapolation.

We show an example sparse-modeling analysis using this standard formula in Appendix~\ref{sec:All-tau-analysis}, where 
 the correlation function is the smeared correlation function at a finite flow-time~\footnote{We put the numerical code and the data on the arXiv page of this paper.}.
Because of the over-smearing problem in the smeared correlation function, we find that the standard formula does not work for the smeared correlation function.

\subsection{Shear viscosity at a finite flow-time}\label{sec:tech-step}
We actually have a smeared correlation function $C(t,\tau)$ instead of $C(\tau)$.
The short $\tau$-regime of $C(t,\tau)$ suffers from the over-smearing correction, and then we have to eliminate these data whose slope is different from the non-smeared ones.
On the other hand, as we will see later (the right panel in Fig.~\ref{fig:Ctau-ft-deps}), the slope of $C(t,\tau)$ in the fiducial $\tau$-regime is universal and does not depend on the flow-time.
Then, we consider the kernel for $C(t,\tau)$ in the fiducial $\tau$-regime can be the same functional form with the one for the non-smeared correlation function.
Thus, the spectral function bears the flow-time dependence of $C(t,\tau)$ while the (reduced) kernel is universal.

Therefore, the integration equation for $C(t,\tau)$ turns to
\beq
C(t, \tau') = N_\tau ^4 \Lambda \int_{-1}^1 d \hat{\omega}' \tilde{\rho}(t, \hat{\omega}') \frac{\cosh [\frac{\omega ' \Lambda }{2} (2 \tau' -1)]}{\cosh (\omega ' \Lambda /2)}.
\eeq
where $\tau'$ runs to the site in the fiducial regime.
From the spectral function in the integral equation above, we will obtain the shear viscosity
\beq
\eta (t,T) = \pi {N_\tau} \tilde{\rho} (t,\omega'=0),\label{eq:eta-t}
\eeq
which depends on the flow-time.
Furthermore, it depends on the lattice spacing, so that we will take the double limits, $a\rightarrow 0$ and then $t \rightarrow 0$ limits, for the obtained $\eta(t,T)$ at a fixed $T$.
After these processes, we will find $\eta(T)/T^3$ at the temperature.

Applying the  sparse modeling method to the smeared correlation function needs three modifications to the standard formula:
Firstly, the input data $C(t,\tau)$ should be restricted only in the fiducial $\tau$-regime as same as the calculation of the one-point function of EMT in Ref.~\cite{Asakawa:2013laa}.
Simultaneously, the kernel matrix is reduced to $K^{red}(\tau'_i,\omega'_j)$, where $\tau'_i$ runs the site only in the fiducial $\tau$-regime.
Secondly, we remove the sum rule given in Eq.~(\ref{eq:sum-rule-Sp}) in the cost function, since the sum rule is related to the correlation function at $\tau=0$ but $C(t,\tau=0)$ is always contaminated by the over-smearing at  finite flow-time.
The third one is the modification of the normalization factor $C(\tau'_0)$ in Eq.~(\ref{eq:norm-C}) to a different value. This modification is not so essential but useful.
Here, we utilize $C(t,\tau'_{ini})$ instead of $C(\tau'_0)$, where $\tau_{ini}$ is the smallest number of site in the fiducial $\tau$-regime.

The modified strategy to obtain the shear viscosity using the gradient flow and the sparse modeling methods can be summarized as follows:

{\bf Step 1':} Measure the correlation function $C(t,\tau/a)$ at  flow-time $t$ using the gradient flow method, and then normalize it using the value of $C(t,\tau/a=\tau_{ini})$.
\beq
C_{calc} (t,\tau'_i) &=& \frac{C(t,\tau'_i)}{N_\tau^4 \Lambda} \left[  \frac{C(t,\tau'_{ini})}{N_\tau^4 \Lambda  \sqrt{\Delta \omega'}}   \right]^{-1}= \frac{C(t,\tau')}{C(t,\tau'_{ini})} \sqrt{\Delta \omega '}.
\eeq

{\bf Step 2':} Construct the reduced kernel
\beq
K^{red}(\tau'_i,\omega'_j)  \equiv \frac{\cosh [\frac{\omega_j ' \Lambda }{2} (2 \tau'_i -1)]}{\cosh (\omega_j ' \Lambda /2)} \sqrt{\Delta \omega'}\label{eq:kernel-SVD}
\eeq
and carry out SVD decomposition.
Here, $\tau'_i$ runs only in the fiducial range for each flow-time, namely $\sqrt{8t}/a \le \tau_i$ and $\sqrt{8t} \le N_\tau- \tau_i$. Here, the second inequality comes from the periodic boundary condition. 
For simplicity, we will drop the second inequality in the present draft and only show the first one. 
Note that we assume that the functional form of the kernel does not change for the reduced data of the smeared correlation function.
We will see that the numerical data of  $C(t,\tau)$ in the fiducial $\tau$-regime support this assumption, and actually this reduced kernel works well in our analyses.

{\bf Step 3':} Estimate an optimal value of $\lambda$. This is the same process as the standard formula.

{\bf Step 4':} Carry out the sparse modeling {\it without the sum rule} (set $\nu=0$ in Eq.~(\ref{eq:App-cost})).
We obtain the spectral function at a finite flow-time from the output of the sparse modeling analysis ($\tilde{\rho}_{calc}$) as
\beq
\tilde{\rho} (t,\omega')= \frac{C(\tau_{ini})}{N_\tau^4 \Lambda \Delta \omega '} \tilde{\rho}_{calc} (t,\omega_j ').\label{eq:tilde-rho-calc-t}
\eeq
We reconstruct the correlation function using the obtained $\tilde{\rho}_{calc}$,
\beq
C_{output}(t,\tau') = \frac{C(t,\tau'_{ini})}{\sqrt{\Delta \omega}} \sum_j K^{red}(\tau'_i, \omega'_j) \tilde{\rho}_{calc} (t,\omega'_j) .
\eeq
Then, we can check the feasibility of the analysis whether  $C_{output} (t,\tau/a)$ reproduces the input correlation function $C(t,\tau/a)$.

{\bf Step 5':} 
Calculate the shear viscosity ($\eta(t, T)/T^3$) on the lattice from the obtained spectral function using Eqs.~(\ref{eq:eta-t}) and (\ref{eq:tilde-rho-calc-t}).
Carry out the same calculations from {\bf Step 1'} to {\bf Step 4'} using the other lattice spacings with keeping a temperature, and then obtain $\eta(t,T)/T^3$ after taking the continuum extrapolation.
Finally, we have $\eta(T)/T^3$ after taking $t \rightarrow 0$ limit.
The demonstration of these processes is out of the present work, and here our goal is to find a stable $\tilde{\rho}(t,\omega')$ from the smeared correlation function.

\section{Simulation setup}\label{sec:setup}
We consider the Wilson plaquette gauge action under the periodic boundary condition at $\beta=6.93$ on $N_s^3 \times N_\tau = 64^3 \times 16$ lattices.
The lattice parameter realizes the system at $T=1.65T_c$. 
We use the relation between $a/r_0$, where $r_0$ denotes the Sommer scale, and $\beta$ in Ref.~\cite{Guagnelli:1998ud}.
Then,  $T/T_c$ is fixed by the resultant values of $Tr_0 = (N_\tau (a/r_0))^{-1}$ using the result at $\beta= 6.20$ in Ref.~\cite{Boyd:1996bx}.
 The gradient flow method in Ref.~\cite{Asakawa:2013laa} gives the thermal entropy, $s/T^3=4.98(24)$, after $a \rightarrow 0$ and then $t\rightarrow 0$ extrapolations.

Gauge configurations are generated by the pseudo-heatbath algorithm with the over-relaxation.
We call one pseudo-heatbath update sweep plus several over-relaxation sweeps
as a ``Sweep". 
To eliminate the autocorrelation, we take $200$ Sweeps between measurements. 
The number of gauge configurations for the measurements is $2,000$. 
Note that it is quite a small number of configurations in comparison with $800,000$ in Ref.~\cite{Nakamura:2004sy} and $6$ million in Ref.~\cite{Pasztor:2018yae}.

Judging from such a poor statistic, we give up a serious estimation of the statistical error and focus on the feasibility of the sparse modeling method to obtain the shear viscosity combing with the gradient flow method.
In \S.~\ref{sec:SpM} we explain how to estimate the spectral function using the central value of the correlation function.
Then, in \S.~\ref{sec:error}, we discuss the systematic and statistical errors in the analysis.
We demonstrate the bootstrap analyses to estimate  statistical uncertainty.

The flowed gauge field is obtained by solving the ordinary first-order differential equation (Eq.~(\ref{eq:gflow})). 
Numerically, we utilize the third-order Runge-Kutta method in which the error per step ($t \rightarrow t+ \varepsilon$) is ${\cal O}(\varepsilon^5)$. 
We take $\varepsilon= 0.01$, and confirm that the accumulation errors are sufficiently smaller than the statistical errors.
The gauge action of the flow is the Wilson plaquette gauge action.

\section{Results of the sparse modeling method for the central value of $C(t,\tau/a)$}\label{sec:SpM}
Now, we demonstrate the sparse modeling analysis using the central value of the measured correlation functions.
Table~\ref{table:parameter} shows the technical parameters in the analysis in this section.
\begin{table}[h]
\begin{center}
\begin{tabular}{|c|c|c|c|c|c|}
\hline
 $s_{cut}$ & $a \omega_{cut}$  & $N_\omega$  & $[\lambda_{min},\lambda_{max}]$ & $N_\lambda$  &  $(\mu,\mu')$  \\
 \hline
 $10^{-10}$ & $ 4.0 $  & $3001$  &  $ [10^{-15} , 10^{2}]$  & $100$  &  $(1.0,1.0)$  \\
 \hline
\end{tabular}
\caption{ Parameters for the present sparse-modeling analysis } \label{table:parameter}
\end{center}
\end{table}
The reason why we take these values of $s_{cut}$ and $a \omega_{cut}$ will be explained below and \S.~\ref{sec:omega-deps}.
Fixing $[\lambda_{min},\lambda_{max}]$, $N_\lambda$, and $(\mu,\mu')$ are correlated with each other. 
This set is a possible choice to find an optimal value of $\lambda$.

\begin{figure}[h]
\begin{center}
\includegraphics[scale=0.5]{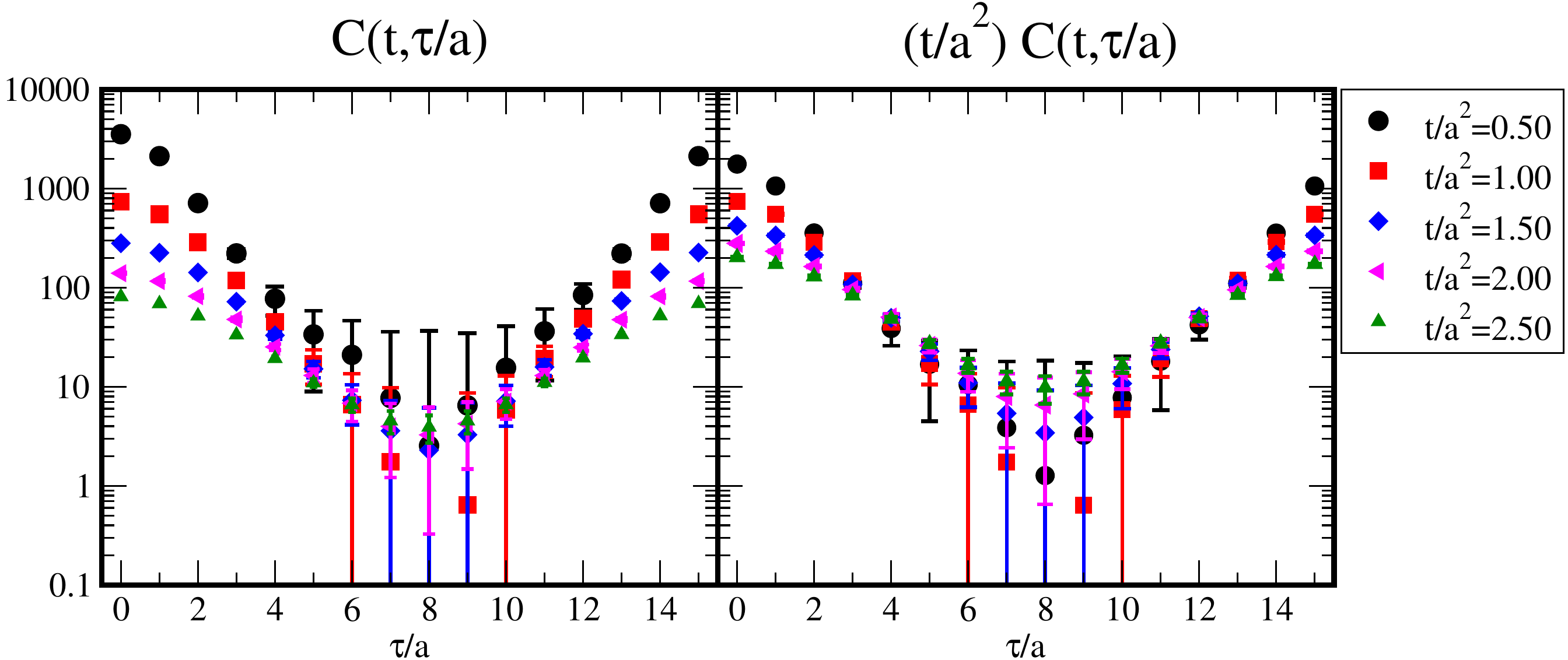}
\caption{The correlation function of $T_{12}^R$ operator at several flow-time.  }
\label{fig:Ctau-ft-deps}
\end{center}
\end{figure}

In {\bf Step 1'} listed the end of \S.~\ref{sec:tech-step}, we measure the correlation function using the gradient flow method in the range of the flow-times $0.50 \le t/a^2 \le 2.50$ with the interval $\Delta t/a^2=0.10$.
Take $t/a^2 = 0.50,1.00,1.50,2.00$, and $2.50$ for example in this section.
The data of $C(t,\tau/a)$ appears in Fig.~\ref{fig:Ctau-ft-deps}.
We see from Fig.~\ref{fig:Ctau-ft-deps} that the longer flow-time data have a smaller statistical error.
Actually, in these poor statistics, namely $2,000$ configurations, the correlation function of the non-smeared $U_{12}$ operator is quite noisy and its central values at some $\tau/a$ often take negative values because of the large fluctuations.
However, all flowed data shown in the left panel in Fig.~\ref{fig:Ctau-ft-deps} (except for only one data-point at $\tau/a=8$ at $t/a^2=1.00$) indicate correctly positive values.
It is a great advantage of the usage of the gradient flow method to obtain the correlation function.

On the other hand, there is a demerit of the usage of the gradient flow for $C(t,\tau/a)$.
Thus, the smearing changes the slope in the short $\tau$-regime because of the over-smearing.
To see the merit and demerit clearly, we rescale the correlation function by multiplying the flow-time, $(t/a^2)C(t,\tau/a)$ (right panel in  Fig.~\ref{fig:Ctau-ft-deps}).
In $\tau/a \lesssim 3$, the slope of the correlation function strongly depends on the flow-time.
It indicates that the kernel term, which is proportional to hyperbolic cosine function, is not available for the whole $\tau$-regime in the flowed correlation function (see also Appendix~\ref{sec:All-tau-analysis}).
On the other hand, around $\tau/a = N_\tau/2 $ the curve of the correlation function does not change, while the statistical errors are reduced thanks to the smearing.
A similar property of the smearing has been reported in the case of the APE smearing for the Wilson loop~(see Fig.~$5$ in Ref.\cite{Bilgici:2009kh}).
To take only the merits of the gradient flow, it is better to take the fiducial $\tau$-regime at  finite flow-time, $\sqrt{8t} < \tau$, for the estimation of $\rho(\omega)$.
Here, we  fix the $\tau$-regime as $3 \le \tau/a \le 13 $ and investigate the flow dependence of the results.

\begin{figure}[h]
\begin{center}
\includegraphics[scale=0.5]{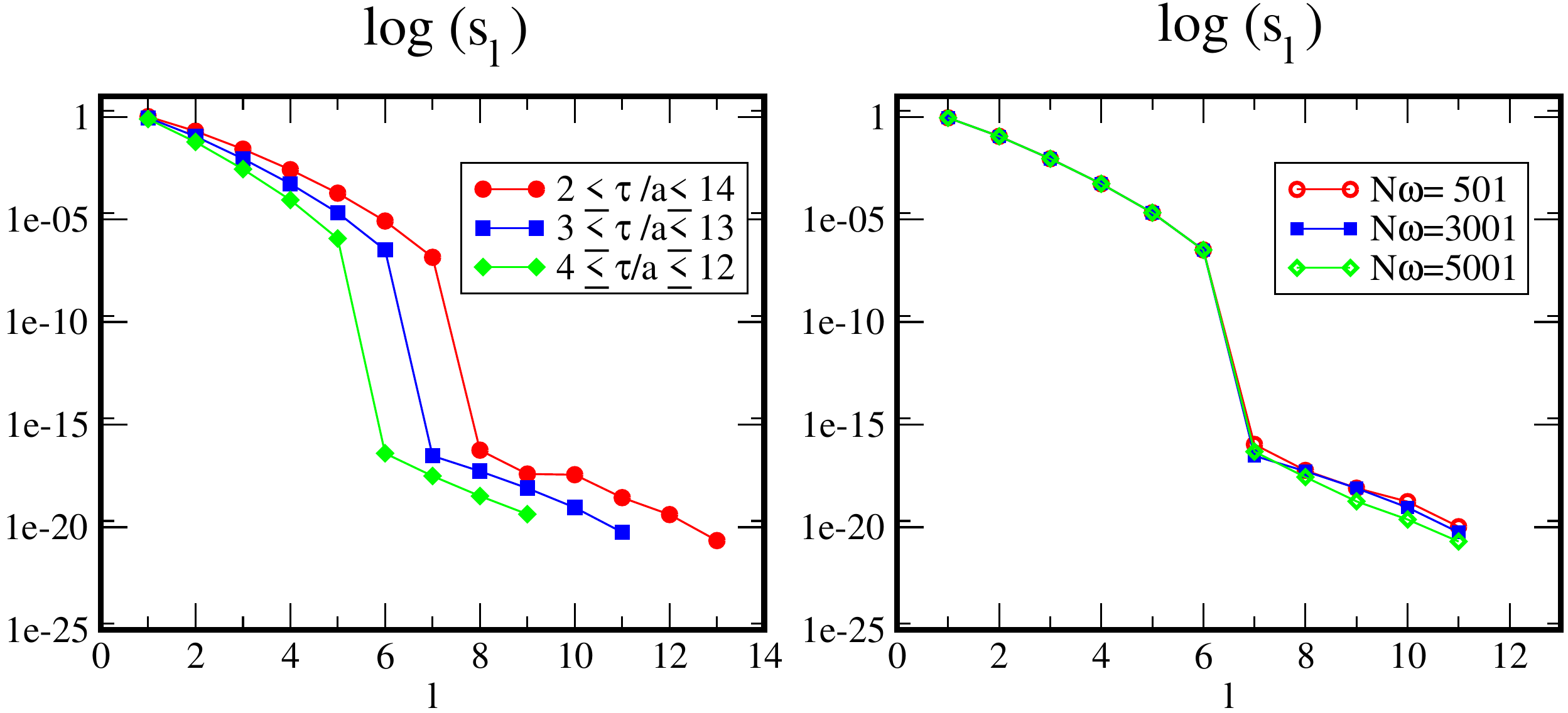}
\caption{$\tau$-regime (left) and $N_\omega$ (right) dependence of the singular of the kernel matrix. Here, we utilize $a \omega_{cut}=4.0$.  }
\label{fig:log-sl-tau-deps}
\end{center}
\end{figure}

The next step, {\bf Step 2'},  is the SVD decomposition of the kernel matrix.
We introduce the cutoff $\hat{\omega}_{cut}=4.0$ and discretize $a \omega$ in the range of $-4.0 \le a\omega \le 4.0$ into $N_{\omega}=3001$ data.
The singular values of the SVD decomposition appear in the left panel of Fig.~\ref{fig:log-sl-tau-deps}.
Note that the vertical axis takes a log scale.
The figure tells us that a large hierarchy more than $10^{10}$ exists between $6$-th and $7$-th singular values if we take $3 \le \tau/a \le 13$. 
Furthermore, we find that the number of $s_l$ larger than $10^{-16}$ is the same with the number of independent $\hat{\tau}_i$ under the periodic boundary condition.
Refer to several condensed matter studies~\cite{Chikano2018a,Chikano2018b,Li2019}, it will saturate to a specific value even though the number of the data points for $C(\tau)$ increases.
It indicates that the number of $\tau_i$, here at most $7$ in taking $2 \le \tau/a \le 12$ under the periodic boundary condition, to analyze this type of kernels is very sparse~\footnote{
We can also estimate a lattice size in the temporal direction to fully analyze the kernel (see Appendix~\ref{sec:App-sl}).}.
To minimize the square error in Eq.~(\ref{eq:chi-sq-SVD}), the contributions of $s_l \rho'_l$ with a sufficiently small $s_l$ are negligible.
Therefore, we introduce the threshold of  the singular value $s_{cut}$ as $s_{cut}=10^{-10}$ in the present analysis.

The most important point is that this property of the singular values is determined only by the kernel matrix Eq.~(\ref{eq:kernel-SVD}) and is independent of the simulation details.
Here, we just assume that the correlation function can be described by a hyperbolic cosine function.

The right panel in Fig.~\ref{fig:log-sl-tau-deps} depicts the $N_\omega$ dependence of the singular values using $3 \le \tau/a \le 13$.
We find that $s_l$ is almost independent of $N_\omega$ if we include $\sqrt{\Delta \omega'}$ in the kernel matrix as Eq.~(\ref{eq:kernel-SVD}).

It is worth to see the correlation function in the IR basis, 
\beq
C'_l = \sum_{\hat{\tau}=3}^{13} U^{t}(l,\hat{\tau}) C(\hat{\tau}) .
\eeq
\begin{figure}[h]
\begin{center}
\includegraphics[scale=0.5]{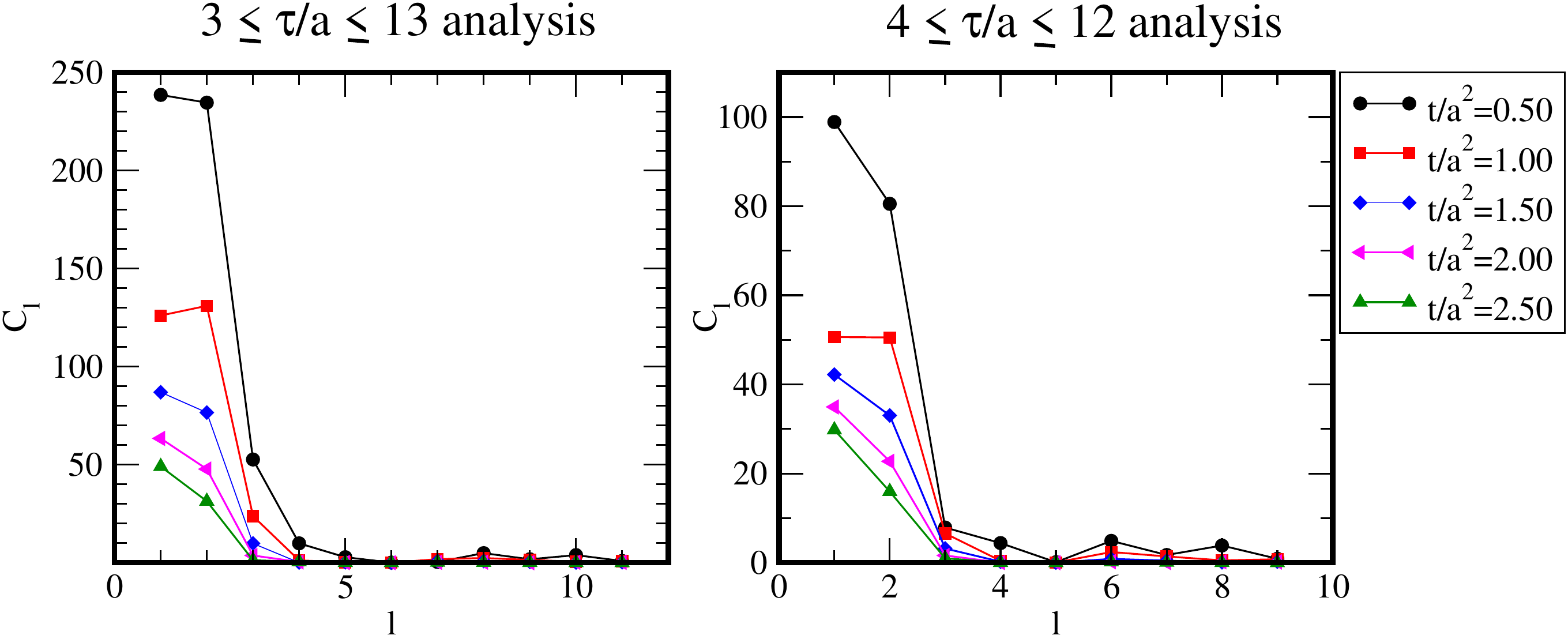}
\caption{The input correlation function in the IR basis ($C_l$) as a function of the label $l$.   }
\label{fig:Cl-ft-deps}
\end{center}
\end{figure}
Figures~\ref{fig:Cl-ft-deps} depicts $C_l$ as a function of $l$.
In comparison with the right panel where $4 \le \tau/a \le 12$, the values of $C_l$ is large.
We take  $3 \le \tau/a \le 13$ and will discuss the results using  $4 \le \tau/a \le 12$ in \S.~\ref{sec:tau-deps}.
We see that $C_l$ with $l \gtrsim 5$ is also sufficiently small, so that it is consistent with the truncation of $s_l$ for $l \ge 7$.

{\bf Step 3'} is the $\lambda$ optimization.
The $\lambda$ dependence of the square error (Eq.~(\ref{eq:chi-sq})) is a monotonically increasing function of $\lambda$~\cite{Shinaoka2017b}.
The optimized $\lambda$ is given by $\lambda^{opt}=1.1 \times 10^{-7}$ -- $1.9 \times 10^{-6}$ in our analysis.
To find this, we scan the value of $\lambda$ in the range of $ 10^{-15} \le \lambda \le 10^{2}$ by changing its exponent with $1 /N_\lambda$ interval.
Once we obtain $\lambda^{opt}$, then we check  whether it shows roughly an inverse-rescaling with $(\mu,\mu')$.

\begin{figure}[h]
\begin{center}
\includegraphics[scale=0.5]{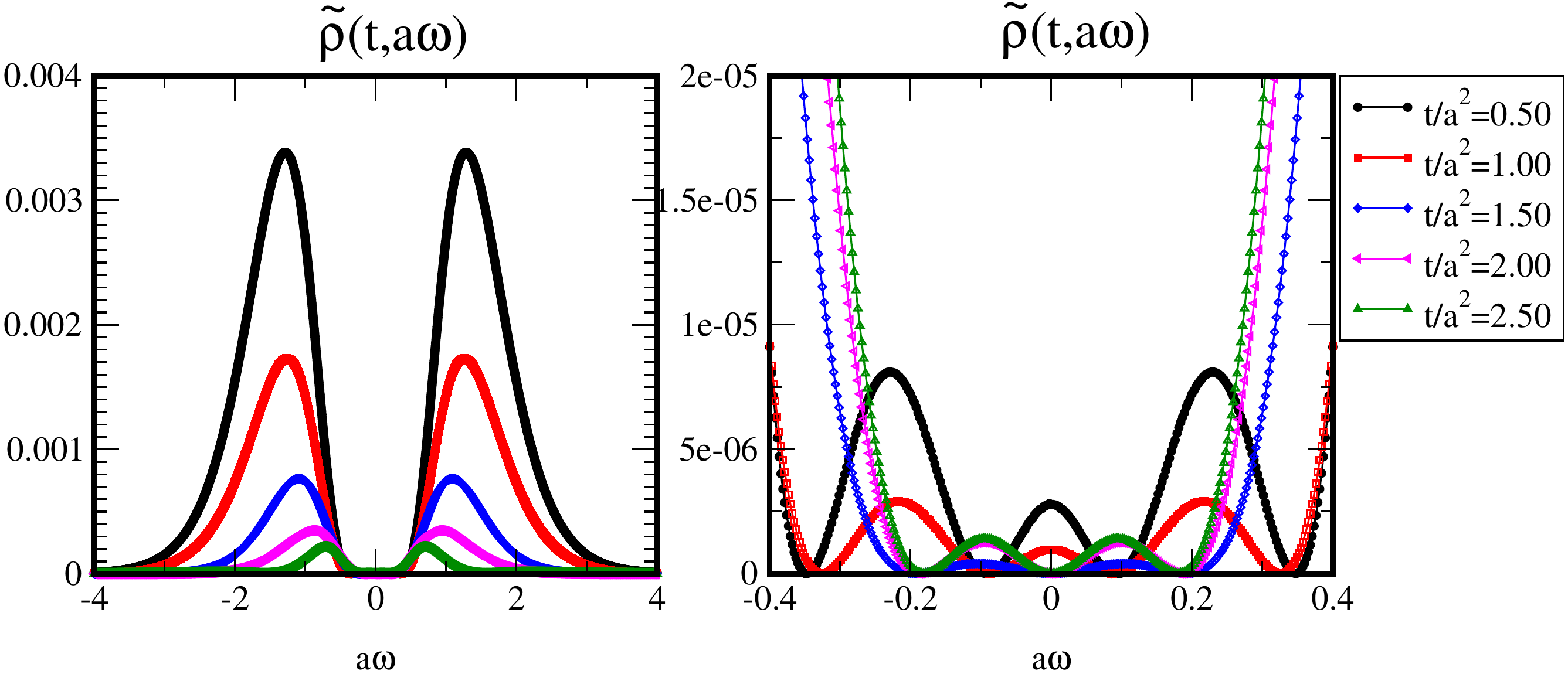}
\caption{The obtained spectral function $\tilde{\rho}(t, a\omega)$ as a function of $a \omega$.  The right panel is an enlarged plot in $\omega \approx 0$ regime.   }
\label{fig:rho-center}
\end{center}
\end{figure}

The results of the spectral function are shown in Fig.~\ref{fig:rho-center}.
First of all, we find the tails of spectral function for all flow-times approach to zero.
It tells us that  $a \omega_{cut}=4.0$ is a reasonable choice in this analysis.
Secondly, we see the integration of $\tilde{\rho}(t,a\omega)$ in terms of $a \omega$, 
\beq
{\mathcal N}=\int_{-a\omega_{cut}}^{a\omega_{cut}} \tilde{\rho}(t,a\omega) d (a\omega),\label{eq:N}
\eeq
monotonically decreases as a function of flow-time.
In more detail, the spectral functions in large $|\omega|$ regime are highly suppressed in the longer flow-time.
The gradient flow gradually reduces the degree of freedom with high-frequency and can be interpreted as a renormalization group flow.
We can see that the results of the sparse modeling analysis give a good account of such an intuitive picture.

The left panel of Fig.~\ref{fig:rho-center} is an enlarged plot, which focuses on $\omega \approx 0$.
The curvatures of $\tilde{\rho}(t,a \omega)$ at $\omega=0$ has an opposite sign between  $t/a^2 =0.50,1.00$ and $t/a^2=1.50,2.00,2.50$.
It remind us that the gradient flow smears the data in $\tau/a < \sqrt{8t}/a$, then the data at $\tau/a =3$ (and $13$) are over-smeared in $t/a^2 \le 1.12$.
Thus, the results in $t/a^2=0.50,1.00$ may suffer from the over-smearing corrections.
To conclude this, we have to investigate the statistical uncertainty of the input $C(t,\tau/a)$.
We will discuss this point in \S.~\ref{sec:error}.

\begin{figure}[h]
\begin{center}
\includegraphics[scale=0.35]{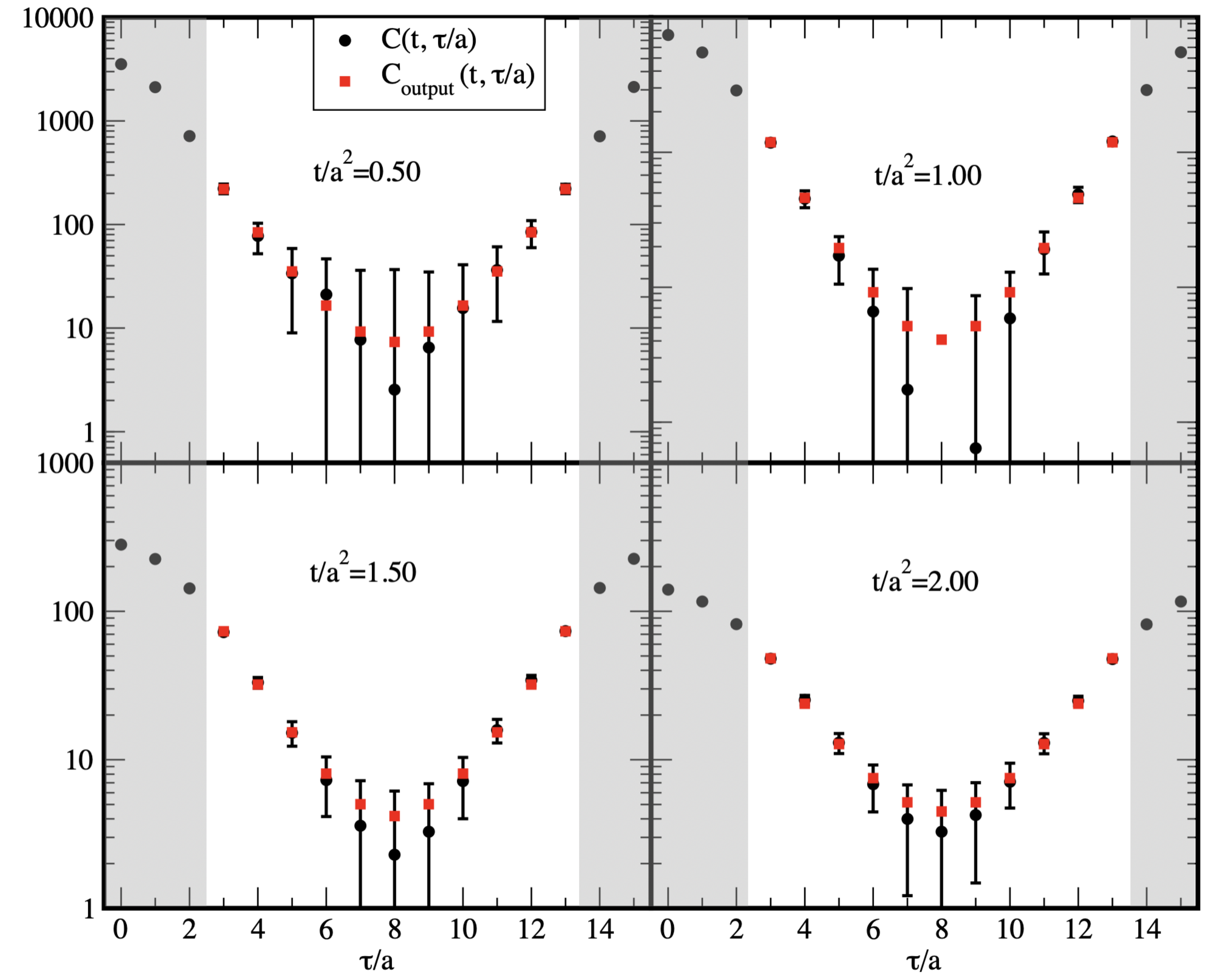}
\caption{Comparison between the input $C(t,\tau/a)$ with the jackknife error and the output $C_{output}(t,\tau/a)$ constructed by the obtained $\tilde{\rho}(t,a\omega)$.  The input $C(t,\tau/a)$ in shadowing regime are not utilized in the analyses. }
\label{fig:comp-Ctau-center-ana}
\end{center}
\end{figure}

Finally, in {\bf Step 4'}, we check whether the obtained spectral function correctly reproduces the input correlation function.
Figure~\ref{fig:comp-Ctau-center-ana} depicts the comparison plot between the input $C(t,\tau/a)$ and the $C_{output}(t,\tau/a)$ constructed by the obtained $\tilde{\rho}(t, a\omega)$.
Note that the input data of the analysis is the central value of $C(t,\tau/a)$, while its jackknife errors are also shown.
We see that most data, especially in the short $\tau$-regime, are consistent between the input and output, while the discrepancies exist around $\tau/a= N_\tau/2$.
We consider that it comes from the large fluctuation of the input data, where some of them are consistent with zero or take a negative value. 
On the other hand, the output data is constructed to be positive, since we utilize the non-negativity condition to find the spectral function.
Actually, the discrepancy around $\tau/aN_\tau=1/2$ in the longer flow-time with less statistical uncertainty becomes smaller.  
Therefore, we believe that the discrepancy will be reduced if we have the input data precisely.

\section{Error estimations}\label{sec:error}
\subsection{ $\omega_{cut}$ dependence}\label{sec:omega-deps}
\begin{figure}[h]
\begin{center}
\includegraphics[scale=0.45]{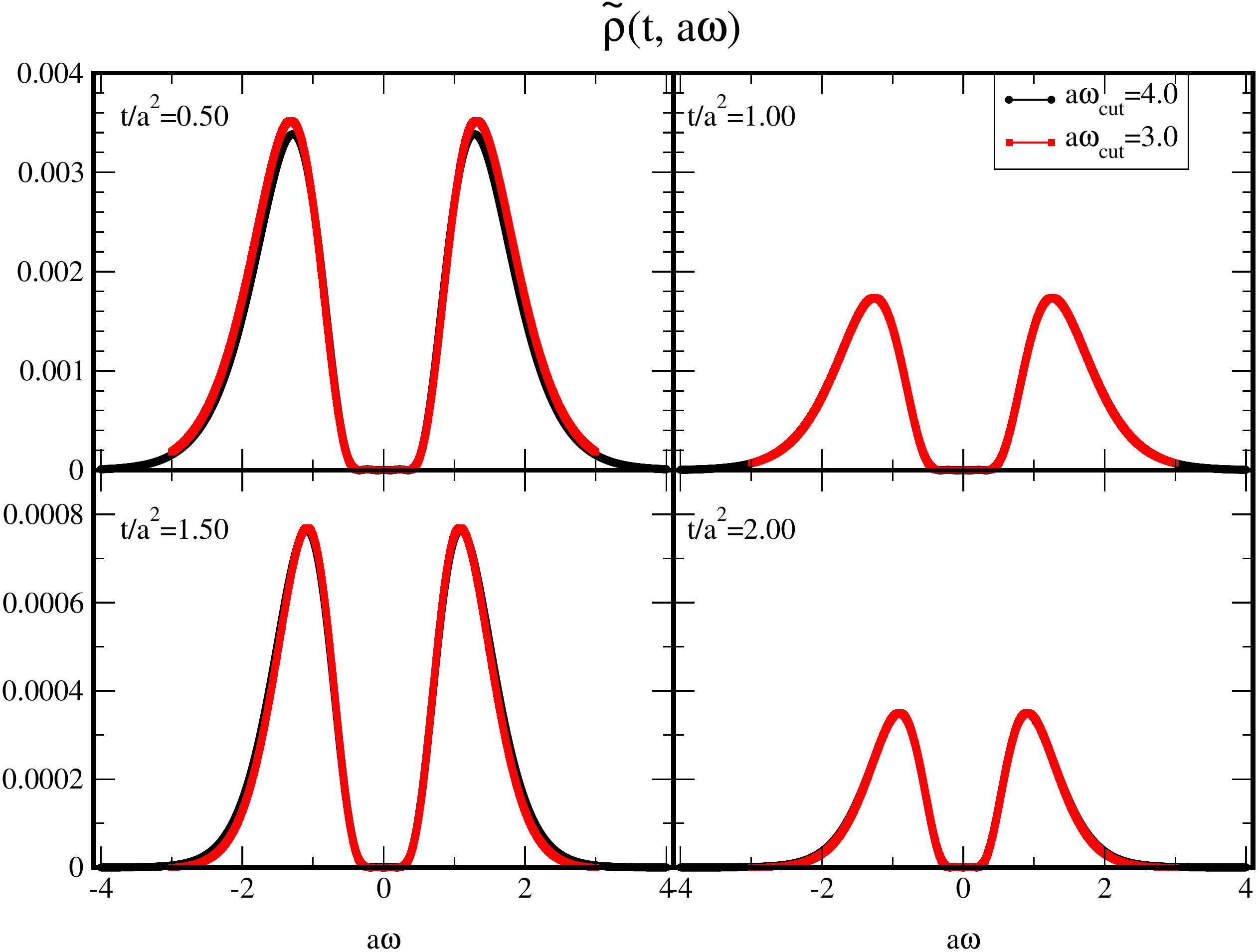}
\caption{The $\omega_{cut}$ dependence of the spectral function for several flow-time data.   }
\label{fig:rho-omegamax-deps}
\end{center}
\end{figure}

Now, we study the systematic uncertainty coming from $\omega_{cut}$, which is artificially introduced.
Figure~\ref{fig:rho-omegamax-deps} shows the comparison of the obtained $\tilde{\rho}(t,a\omega)$ between taking $a\omega_{cut}=3.0$ and $a\omega_{cut}=4.0$ for several flow-times data.
The tails of the spectral functions using $a\omega_{cut}=4.0$ for all flow-times approach to zero, while the ones using $a \omega_{cut}=3.0$ for $t/a^2=0.50, 1.00$ are the middle of the decreasing.
Nevertheless, the shape of spectral function for $t/a^2=0.50, 1.00$ does not depend on the value of $\omega_{cut}$ so much.
It implies the stability of this sparse modeling analysis when the artificial parameter $\omega_{cut}$ is changed.

On the other hand, the tails of $\tilde{\rho}(t,a\omega)$ for $t/a^2=1.50, 2.00$ are closed to zero even though we utilize $a\omega_{cut}=3.0$, since the higher frequency modes are suppressed by the long flow processes.
From this point of view, although the sparse-modeling method can apply to the non-smeared data, the smearing may practically stabilize the analysis with the truncation of $\omega$~{\footnote{Applying the sparse modeling analysis  to mock data whose spectral function has a constant mode has been considered~\cite{Tomiya}. }}.

\subsection{$\tau$-regime dependence and the fiducial window of the gradient flow}\label{sec:tau-deps}
Next, we investigate the relationship between a choice of $\tau$-regime and the gradient flow-time. 
The fiducial $\tau$-regime is theoretically given as $\tau > \sqrt{8t}$, and we expect that  the data does not suffer from an over-smearing.
\begin{figure}[h]
\begin{center}
\includegraphics[scale=0.45]{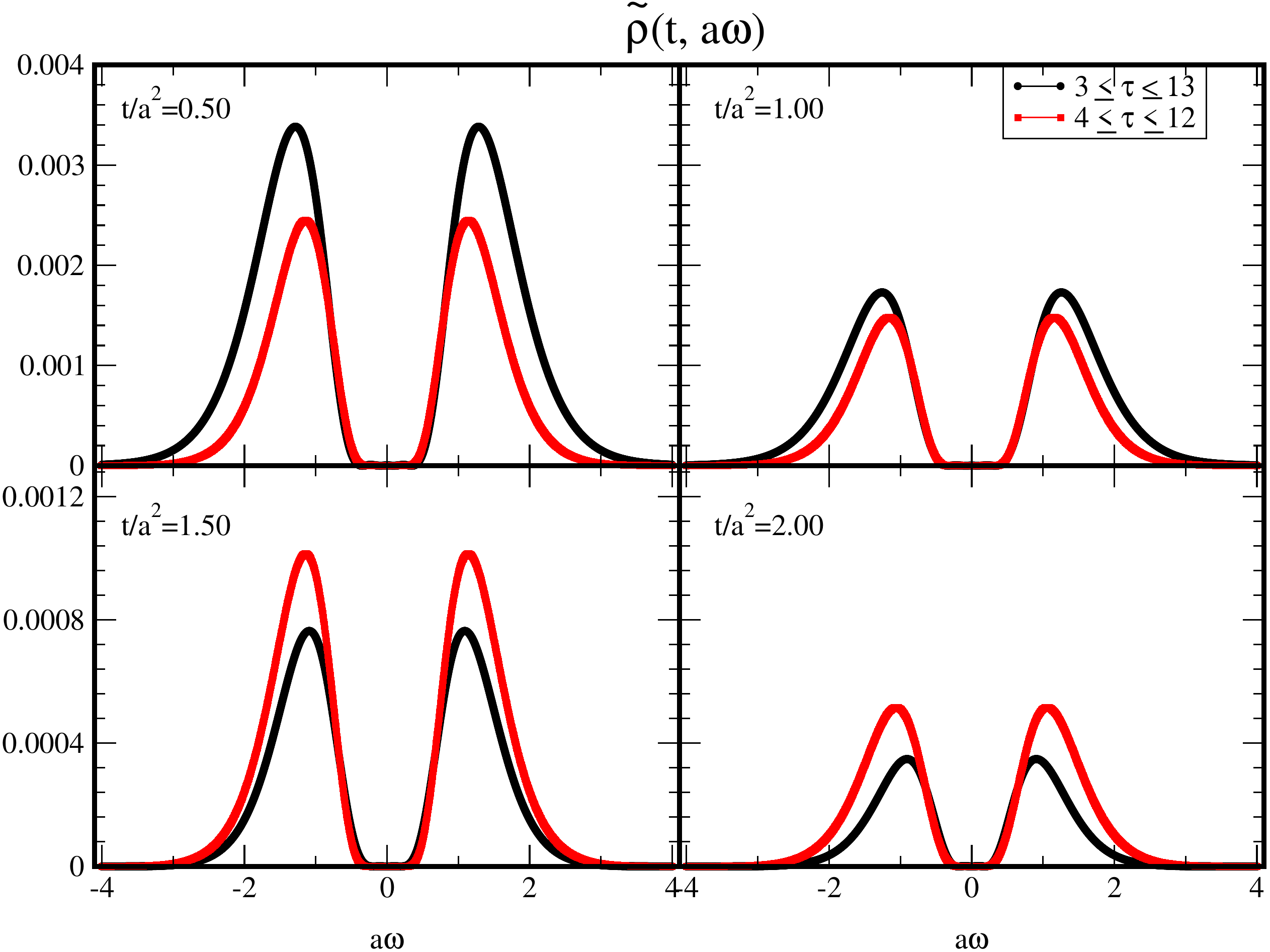}
\caption{The $\tau$-regime dependence of the spectral function for several flow-time data.   }
\label{fig:rho-Ntaudata-deps}
\end{center}
\end{figure}
In other words, if we take $3 \le \tau/a \le 13$, then $C(t,\tau/a)$ in $t/a^2 < 1.12$ are in the fiducial regime, while $\tau/a=3$ (and $13$) must have a strong over-smearing correction in $t/a^2 > 1.12$.
On the other hand, if we take $4 \le \tau/a \le 12$, then $C(t,\tau/a)$ in $t/a^2 < 2.00$ stay in the fiducial regime.

Figure~\ref{fig:rho-Ntaudata-deps} shows the comparison of the spectral functions between the $3 \le \tau/a \le 13$ and $4 \le \tau/a \le 12$ analyses for $t/a^2 =0.50, 1.00, 1.50, 2.00$.
First of all, as we explained, the gradient flow with each fixed $\tau$-regime (same color symbols) reduces $\mathcal{N}$ in Eq.~(\ref{eq:N}).
On the other hand, we naively expected that $\mathcal{N}$ must decrease if we take the narrow regime of $\tau$ since the effective degrees of freedom are truncated.
However, in $t/a^2=1.50, 2.00$, $\mathcal{N}$ of the red data ($4\le \tau/a \le 12$) is larger than the one of the black data ($3 \le \tau/a \le 13$).
The discrepancy is slightly beyond the statistical error, which we will study in \S.~\ref{sec:stat-error}.
It implies that the corrections to the spectral function from over-smearing in the black data are strong rather than the influence on $\tilde{\rho}(t,a\omega)$ from the truncation of $C(t,\tau/a)$.

\begin{figure}[h]
\begin{center}
\includegraphics[scale=0.35]{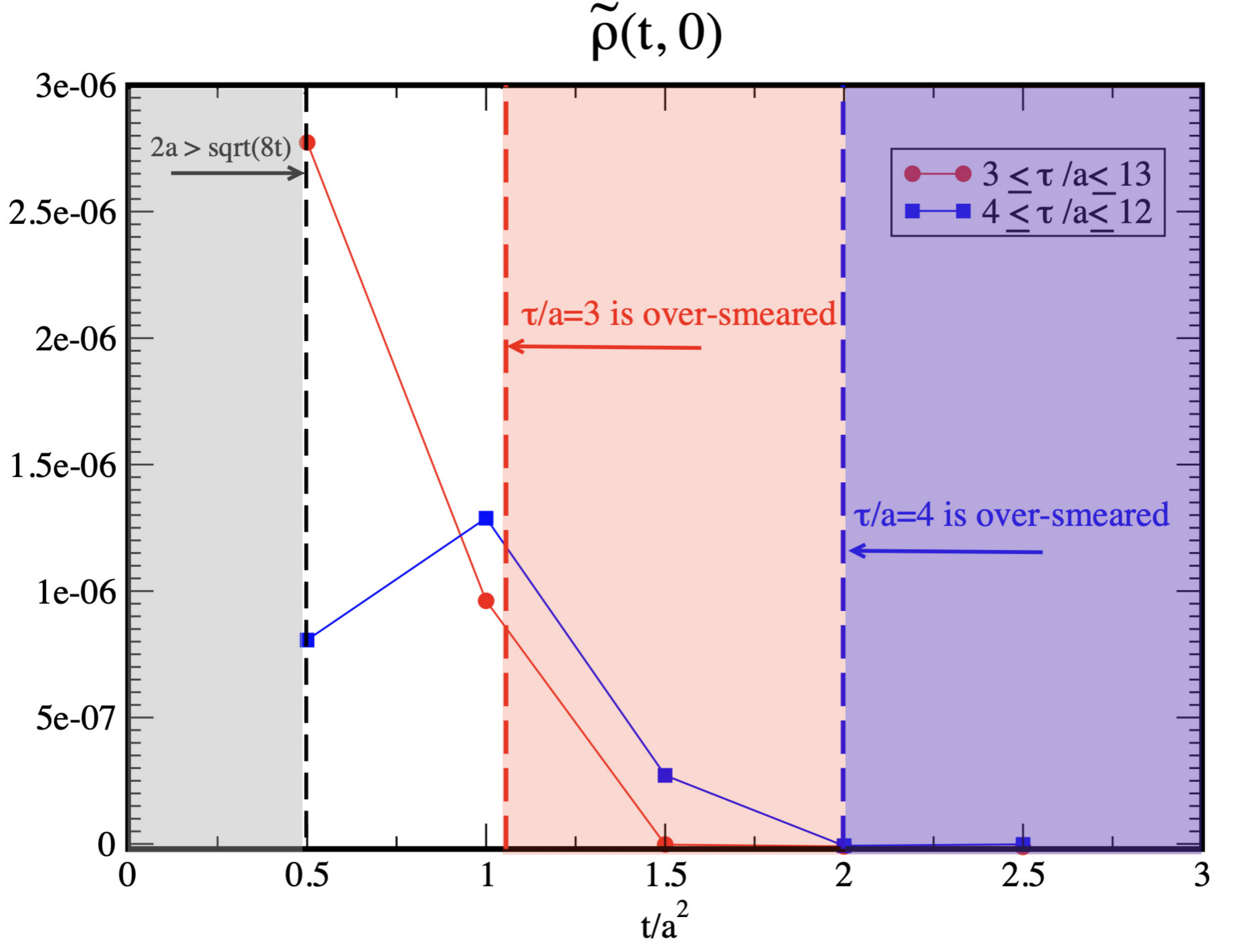}
\caption{The flow-time and $\tau$-regime dependence of $\tilde{\rho}(t,\omega=0)$. The fiducial window ($2a < \sqrt{8t} < \tau$)  is indicated
by the dashed lines. $0.50 < t/a^2 < 1.12$ (non-shadowing) is the fiducial window for the  $3 \le \tau/a \le 13$ analysis, while $0.50 < t/a^2 < 2.00$ (non-shadowing and red-shadowing) depicts the one for the $4 \le \tau/a \le 12$ analysis.   }
\label{fig:rho0-ft-deps}
\end{center}
\end{figure}
Figure~\ref{fig:rho0-ft-deps} shows $\tilde{\rho}(t,0)$ as a function of the flow-time.
Here, we summarize that the fiducial regime of the flow-time for each $\tau$-regime analysis.
The orders of all $\tilde{\rho} (t,0)$ in the fiducial window are the same.
In more detail, the value itself seems to have a discrepancy between two $\tau$-regime analyses. 
We will discuss the difference in the next section after including the statistical uncertainty of $C(t,\tau/a)$ since a tiny noise of the correlation function often leads to a large difference in the spectral function in this kind of the ill-posed inverse problems.

\subsection{Statistical errors }\label{sec:stat-error}
Now, we try to include the statistical uncertainty of the correlation function in our analysis.
The statistical uncertainty of the correlation function ($C(t,\tau/a)$) is not directly related to the error of the spectral function ($\tilde{\rho}(t,a\omega)$) since these two quantities are related to each other through the integration equation.
On the other hand, the correlation function in the IR basis is linearly related to the spectral function in the same basis (Eq.~(\ref{eq:Cl-sl-rhol})).
Thus, we expect that the statistical uncertainties of these quantities satisfy
\beq
\Delta C_l \sim s_l \Delta \rho_l.\label{eq:error-Cl-rhol}
\eeq

We propose the bootstrap method to estimate the statistical error of the spectral function as follows: 
We resample $N_{boot}$ sets of $2,000$ data of $C(t,\tau/a)$ for each configuration, where the overlapping selection of the configurations for one bootstrap sample is allowed.
For each set of the bootstrap sample, we take the average of $2,000$ data of $C(t, \tau/a)$ and carry out the sparse modeling analysis using its mean value.
The statistical errors of the spectral function are calculated by the variance over $N_{boot}$ samples.
It allows finding the asymmetric errors, so that it is a good tool to calculate the errors of $\tilde{\rho}$ which should be non-negative as a constraint.
Here, we take $N_{boot}=1,000$.

\begin{figure}[h]
\begin{center}
\includegraphics[scale=0.45]{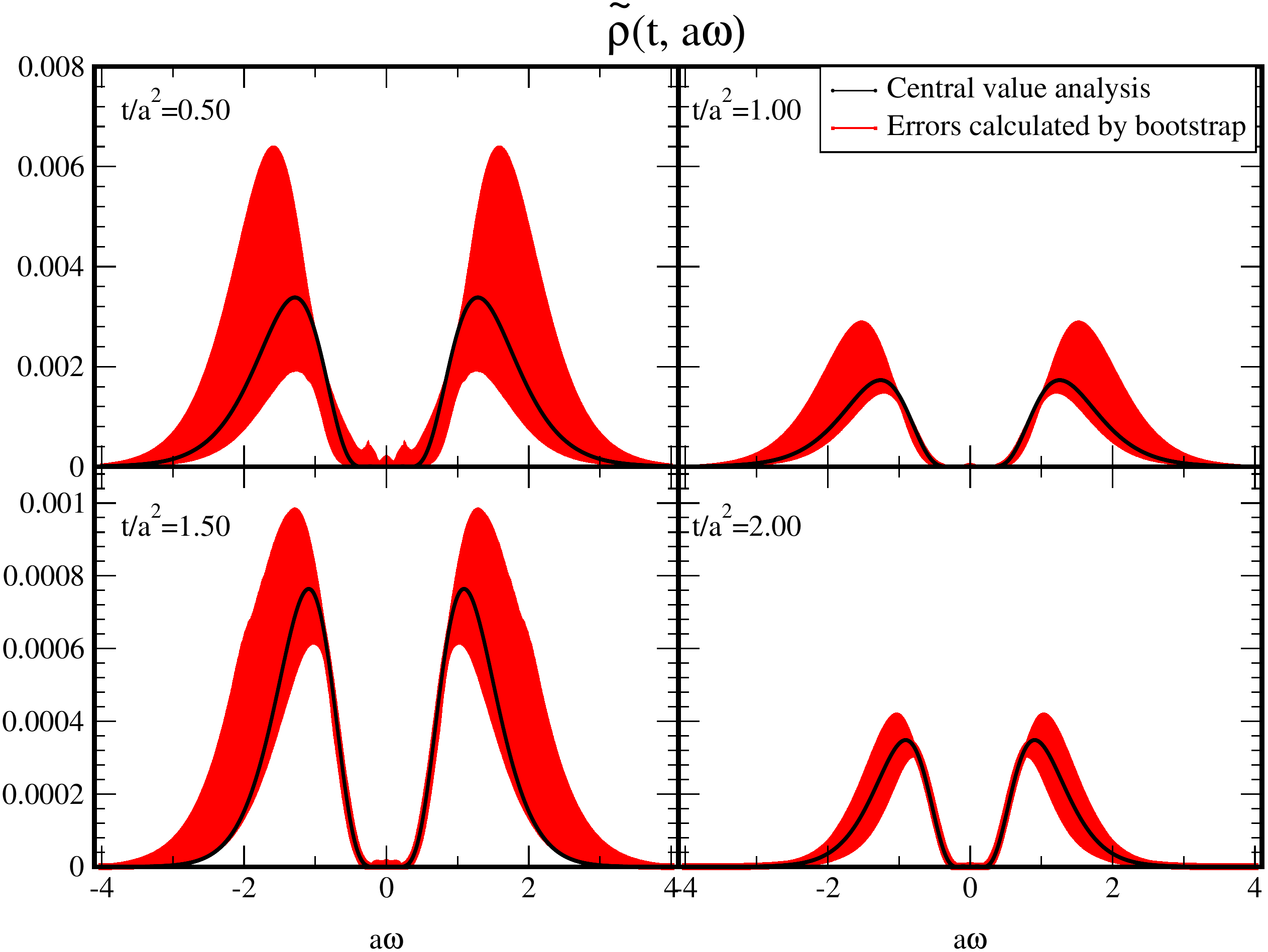}
\caption{The statistical error of the spectral function using the bootstrap analysis. }
\label{fig:rho-errors}
\end{center}
\end{figure}
Figure~\ref{fig:rho-errors} depicts the spectral function with the bootstrap errors.
We also plot the result of the central-value analysis shown in Fig.~\ref{fig:rho-center}, which is inside the error bound.

We check whether the bootstrap errors satisfy an expected relationship Eq.~(\ref{eq:error-Cl-rhol}).
Here, using Eq.~(\ref{eq:rho-C-IR-basis}), $\Delta C_l$ is estimated as the variation of $C_l$  transformed by the bootstrap sample of $C(t,\tau/a)$,
while $\Delta \rho_l$ is done as the variation of the obtained $\tilde{\rho}(t,\hat{\omega})$ for each bootstrap sample.
\begin{figure}[h]
\begin{center}
\includegraphics[scale=0.5]{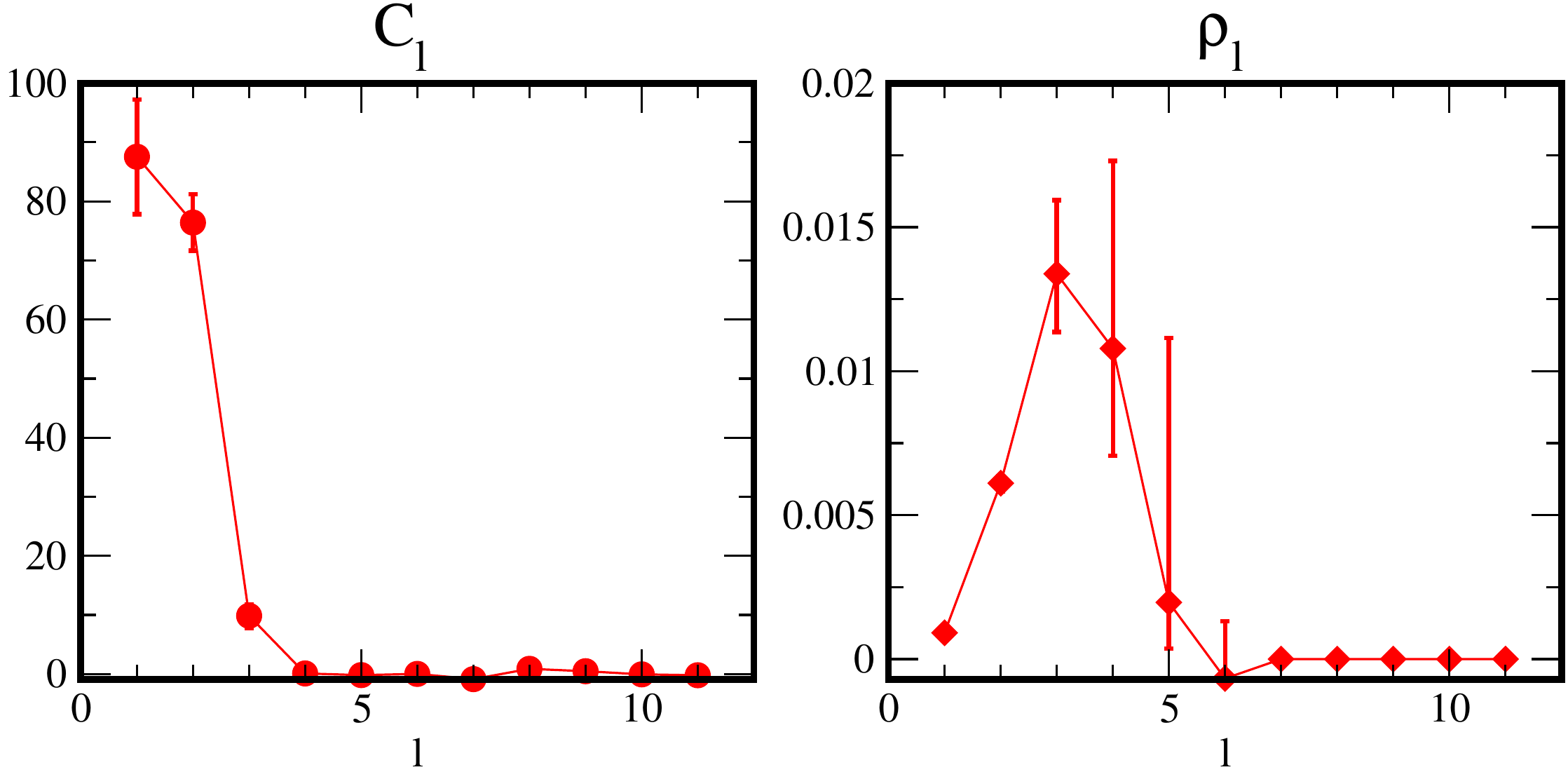}
\caption{The correlation function and the spectral function in the IR basis as a function of $l$.  The flow-time of the data is $t/a^2=1.50$. }
\label{fig:Cl-rhol}
\end{center}
\end{figure}
Figure~\ref{fig:Cl-rhol} shows the $l$-dependence of $C_l$ and $\rho_l$.
We see that the statistical errors of them roughly satisfies $\Delta \rho_l \sim \Delta C_l /s_l $.
Note that $\rho_l$ with $l \ge 7$ are consistent with zero since these modes are eliminated by the cut of the singular values in the analysis.

\begin{figure}[h]
\begin{center}
\includegraphics[scale=0.35]{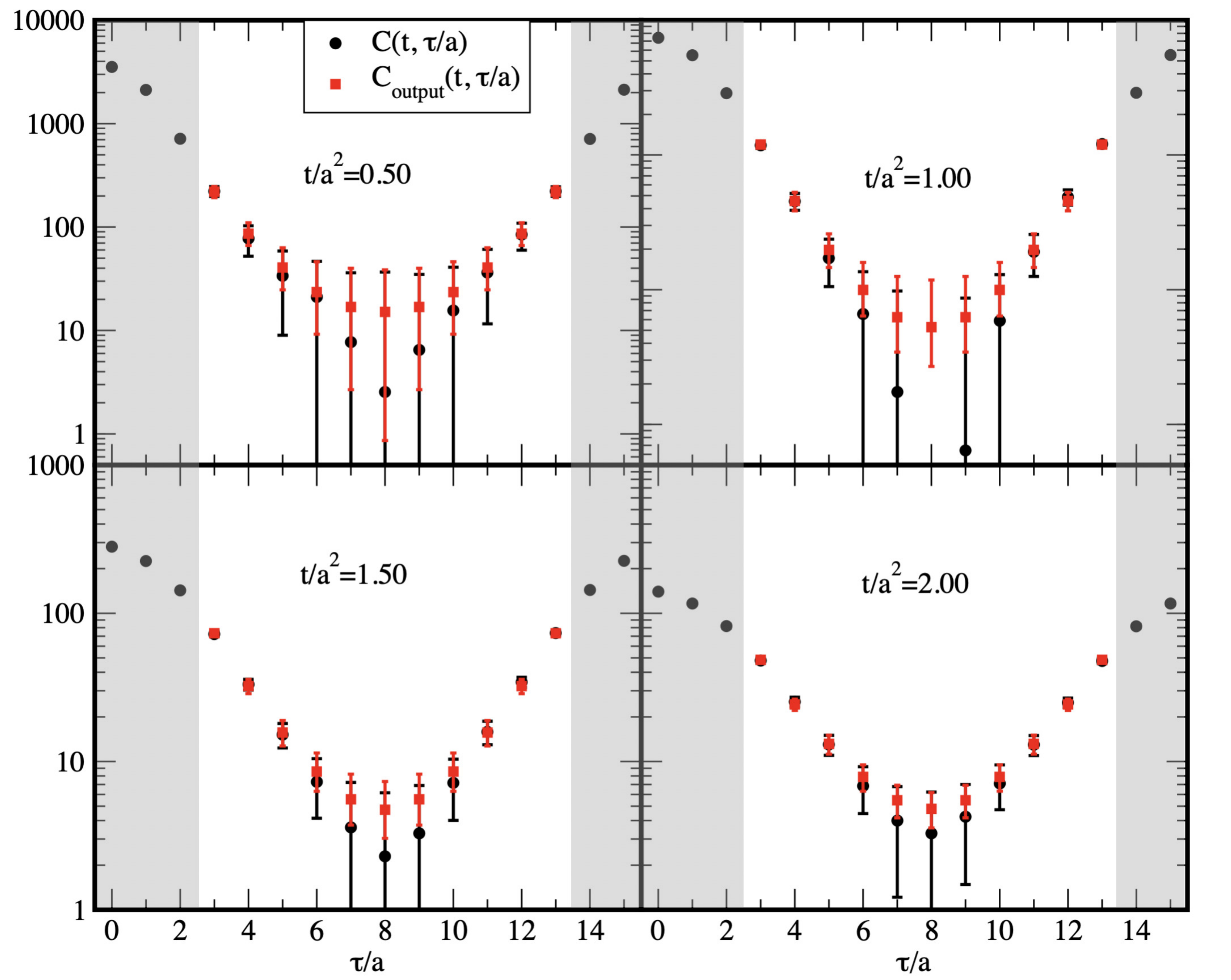}
\caption{The comparison between the input correlation function with the Jackknife error and the output one with the bootstrap error.  The data $C(t,\tau/a)$ in shadowing regime are not utilized in the analyses.}
\label{fig:input-output-Ctau}
\end{center}
\end{figure}

We also show the comparison of the statistical error bars between the input and output $C(t,\tau/a)$ in Fig.~\ref{fig:input-output-Ctau}.
The error bars between them in the short $\tau$-regime are consistent with each other.
The error bars of the output around $\tau/a=N_\tau/2$ are smaller than the ones of the input.
We consider that it comes from the non-negativity condition of the spectral function during the sparse modeling analysis.
It makes $C(t,\tau/a)$ positive correctly, and then the condition reduces the error bars of $C_{output}$.
We can conclude that the bootstrap method is a reasonable estimation method of the statistical errors for the sparse modeling analysis.

\subsection{Flow-time dependence of the shear viscosity}
Finally, we discuss the flow-time dependence of the shear viscosity with the statistical uncertainty.
Since the number of configurations is quite poor, then we cannot give a conclusive value of the shear viscosity.
Furthermore, we first take the continuum limit of $\eta(t,T)$ obtained by the lattice simulation with several different lattice spacings at the fixed temperature. After that, we could discuss the flow-time dependence~\footnote{
Otherwise, physical observables diverge in the inverse-ordered extrapolations, $t\rightarrow 0$ and then $a \rightarrow 0$.}.
Thus, in the present analysis, we do not include the systematic uncertainties coming from the continuum extrapolation.
Therefore, we show the results to see naive flow-time dependence at a fixed lattice spacing.

\begin{figure}[h]
\begin{center}
\includegraphics[scale=0.25]{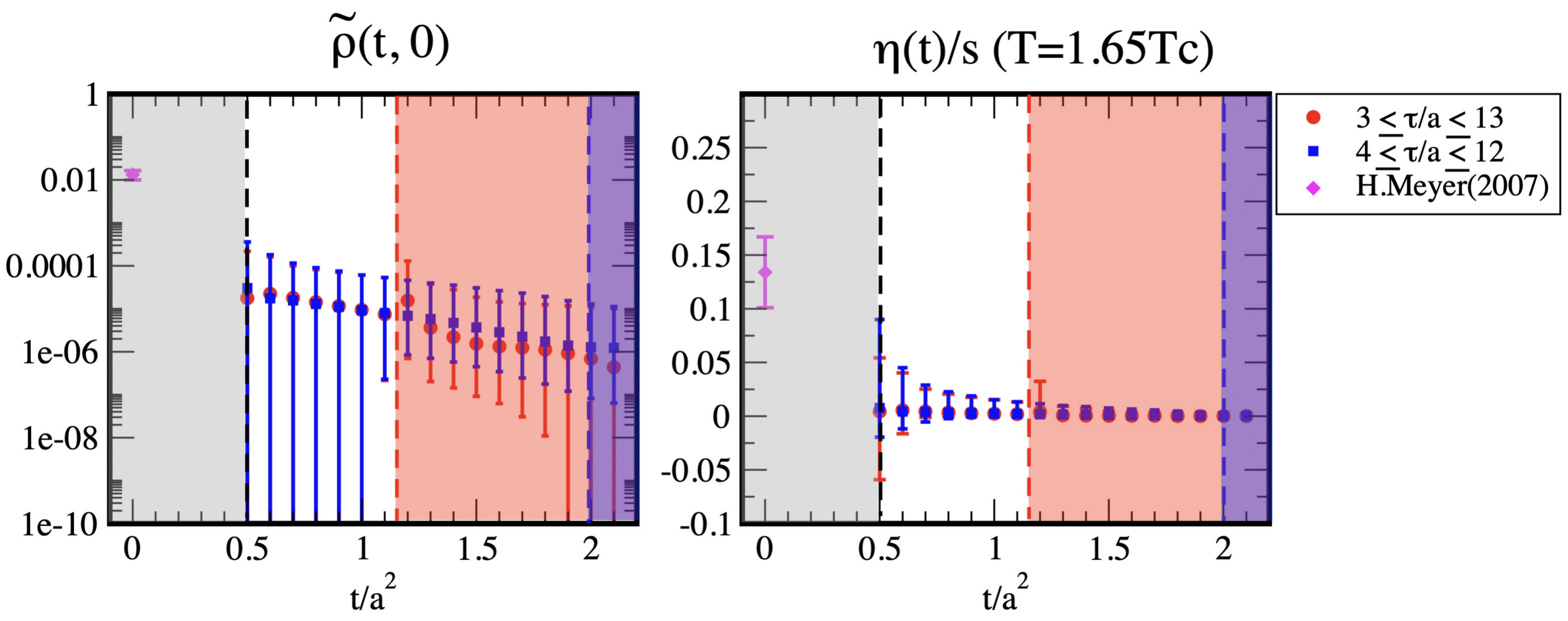}
\caption{ (Left) The spectral function at $\omega=0$. (Right) The ratio between the shear viscosity and the thermal entropy ($\eta/s$). 
Horizontal axes denote the flow-time. Magenta data at $t/a^2=0$ denotes the result by H.~Meyer at the same temperature~\cite{Meyer:2007ic}.
The fiducial window ($2a < \sqrt{8t} < \tau$)  is indicated by the dashed lines, where the meaning of each line and shadowing region is the same with the one in Fig.~\ref{fig:rho0-ft-deps}.   }
\label{fig:rho0-eta}
\end{center}
\end{figure}
The left panel in Fig.~\ref{fig:rho0-eta} depicts the value of $\tilde{\rho}(t,0)$ in the $3 \le \tau/a \le 13$ and $4 \le \tau/a \le 12$ analyses.
We find that both analyses are consistent with each other
in non-shadowing regime ($0.50 < t/a^2 < 1.12$), where all data in both analyses are in the fiducial $\tau$-regime.
However, the discrepancy appears in red-shadowing regime ($1.12 < t/a^2 < 2.00$), where the data at $\tau/a=3$ is over-smeared.
Furthermore, the slope of the flow-time dependence of the $4 \le \tau/a \le 12$ analysis is milder than the one of the $3 \le \tau/a \le13$ analysis.
It tells us the importance of taking the fiducial window in the sparse modeling analysis.

The ratio between the shear viscosity and the thermal entropy is shown in the right panel in Fig.~\ref{fig:rho0-eta}.
Here, we utilize $s/T^3=4.98$ at this temperature, which is given in Ref.~\cite{Asakawa:2013laa}.
We also plot the result in Ref.~\cite{Meyer:2007ic} at $t/a^2=0$ as a reference.
A theoretical large-$N_c$ analysis based on AdS/CFT correspondence for ${\mathcal N}=4$ super Yang-Miils theory gives the lower bound for $\eta/s=1/4\pi$~\cite{Son:2007vk}~\footnote{The $1/N_c$ correction terms to $\eta/s$ has not yet been determined even for its sign in the finite $N_c$~\cite{Kats:2007mq}.}.
Although our result seems to be smaller than the previous works, we can conclude that our result also indicates the small $\eta/s$, which describes the perfect liquid property in the QGP phase.

\section{Summary}\label{sec:summary}
We have proposed the sparse modeling analysis to estimate the spectral function from the smeared correlation functions.
We have described how to obtain the shear viscosity from the correlation function of the renormalized EMT measured by using the gradient flow method for the quenched QCD at  finite temperature.
The gradient flow method reduces the statistical uncertainty of the correlation functions thanks to its smearing property,
while the smearing breaks the sum rule of the spectral function.
 Therefore, we have eliminated the over-smeared data  when we analyze the spectral function.

We have first given the standard formula of the sparse modeling analysis where we assume  $C(t,\tau)$ in the whole $\tau$-regime is available for the analysis.
Then, we have modified the formulation to investigate the smeared correlation function, where the over-smeared data of $C(t,\tau/a)$ are eliminated.
We have shown that the sparse modeling analyses in the IR basis looks stable even using very limited data of the correlation function and the obtained spectral function reproduces the input correlation function.
Several systematic uncertainties of the analysis are well under control. 
We have also demonstrated the bootstrap analysis for estimating the statistical errors.
Although the statistical error of our result is sizable because of poor statistics of our data, we have shown that the bootstrap analysis seems to be promising since the expected relationship of the errors between $C_l$ and $\rho_l$ have been satisfied.

If we will collect $6$ million configurations as the same with the work~\cite{Pasztor:2018yae}, a naive estimation following $1/\sqrt{N_{conf.}}$ would give a few $\%$ relative error for $\eta/s$.
It looks very promising toward the precise determination of the shear viscosity.

\acknowledgments
We are grateful to  T.~Hirano and A.~Tomiya for useful comments.
The numerical simulations were carried out on SX-ACE 
 at Cybermedia Center (CMC) and Research Center for Nuclear Physics (RCNP), 
Osaka University.
We also acknowledge the help of CMC in tuning the gradient-flow code.
This work partially used computational resources of
 HPCI-JHPCN System Research Project (Project ID: jh200031) in Japan.
 This work was supported in part by Grants-in-Aid for Scientific Research 
through Grant Nos.
 15H05855,
and 19K03875
, which were 
provided by the Japan Society for the Promotion of Science (JSPS), and in part 
by the Program for the Strategic Research Foundation at Private Universities 
``Topological Science'' through Grant No.\ S1511006, which was supported by the
Ministry of Education, Culture, Sports, Science and Technology (MEXT) of 
Japan.
The calculations were partially performed by the supercomputing system SGI ICE X at the Japan Atomic Energy Agency.
This work was partially supported by JSPS-KAKENHI Grant Numbers 18K11345.

\appendix
\section{ADMM algorithm}\label{sec:ADMM}
Here, we give a review of the ADMM algorithm, which gives a solution to the optimization problem with several constraints.
Following Refs.~\cite{Shinaoka2017b, ADMM1, ADMM2}, we explain how to build the numerical code in detail to make the paper self-contained.

The problem is to find a minimization of $F(\vec{\rho}')$ in Eq.~(\ref{eq:cost-fn-L1}) with respect to $\vec{\rho}'$ under additional constraints.
Using the conventional notation, we change the variables as $\vec{C} \rightarrow \vec{y}$ and $\vec{\rho} \rightarrow \vec{x}$.
The cost function is written by
\beq
F(\vec{x}') = \frac{1}{2} \| \vec{y}' - S \vec{x}'  \|^2_2 + \lambda \| \vec{x}' \|_1.
\eeq
Here, we consider two constraints:
\beq
x_j \ge 0, ~~~ \langle \vec{x} \rangle \equiv \sum_j x_j = 1.
\eeq
These constraints correspond to the non-negativity of the spectral function and the sum rule, respectively.
Here, $\vec{x} = V \vec{x}'$ using the matrix $V$ obtained by the SVD decomposition, and we have used a convention that vectors with prime denote quantities represented in the IR (SVD) basis.
The dimension of this optimization problem is given by $L = \mbox{min} (N_\tau, N_\omega)$, where $N_\tau$ and $N_\omega$ denotes the size of $\vec{y}$ and $\vec{x}$, respectively.
Actually, we may further reduce $L$ by introducing a cut of the singular value in analysis.

In the ADMM algorithm~\cite{ADMM1, ADMM2}, we introduce auxiliary vectors $\vec{z}$ and $\vec{z}'$, and consider  the minimization of the function
\beq
\tilde{F} (\vec{x}',\vec{z}',\vec{z}) &=& \frac{1}{2\lambda} \| \vec{y}' - S \vec{x}'  \|^2_2 - \nu ( \langle  V \vec{x}' \rangle -1 ) 
+\| \vec{z}' \|_1 + \lim_{\gamma \rightarrow \infty} \sum_j \Theta (-z_j),\label{eq:App-cost}
\eeq
to be
\beq
\vec{z}' = \vec{x}', ~~~~\vec{z} = V \vec{x}'.
\eeq
Here, the sum rule imposed by the Lagrange multiplier $\nu$, and the non-negativity is represented by $\gamma$.
Thus, the auxiliary vectors $\vec{z}'$ and $\vec{z}$ inflect the sum rule and the non-negativity, respectively.
The description to realize the first constraint, $\vec{z}' = \vec{x}'$, is given by two kinds of coefficients; (normalized) Lagrange multipliers $\vec{u}'$ and a coefficient $\mu'$ for $\| \vec{z}' - \vec{x}' \|^2_2$.
The parameter $\mu'$ controls the speed of the convergence, while $\vec{u}'$ is iteratively updated together with its conjugate variables $\vec{z}'$.
Similarly, $\mu$ and $\vec{u}$ are introduced to realize the second constraint, $\vec{z}=V \vec{x}$.

The explicit description is following:
\beq
\vec{x}' &\leftarrow& \left( \frac{1}{\lambda} S^t S + (\mu'+\mu) \mathbf{1}  \right)^{-1} \left( \frac{1}{\lambda} S^t \vec{y}' + \mu' (\vec{z}' -\vec{u}') +\mu V^t (\vec{z}-\vec{u} + \nu V^t \vec{e}) \right) \label{eq:ADMM-nu}\\
&\equiv& \vec{\xi}_1 + \nu \vec{\xi}_2\\
\vec{z}' & \leftarrow & S_{1/\mu'} (\vec{x}' + \vec{u}'),\label{eq:ADMM-Salpha}\\
\vec{u}' &\leftarrow & \vec{u}' + \vec{x}' -\vec{z}' ,\\
\vec{z} & \leftarrow & \mathcal{P}_+ (V\vec{x}' + \vec{u}),\\
\vec{u} & \leftarrow & \vec{u} + V \vec{x}' -\vec{z},\label{eq:ADMM-end}
\eeq
where $e_i =1$ and 
\beq
\nu = \frac{1 - \langle V \vec{\xi}_1 \rangle }{\langle V \vec{\xi}_2 \rangle}.
\eeq
Here, $\mathcal{P}_+$ denotes a projection operator onto non-negative quadrant; $\mathcal{P}_+ z_j = \max(z_j,0)$ for each component.
The explicit form of $S_\alpha$ in Eq.~(\ref{eq:ADMM-Salpha}) is defined by
\beq
S_\alpha (x) =
\left\{ \begin{array}{ll}
x-\alpha & (x>\alpha) \\
0 & (-\alpha \le x \le \alpha) \\
x+\alpha & (x< -\alpha) .\\
\end{array} \right.
\eeq

A simple choice of the initial vectors is a set of zero vectors for all.
The update Eqs.(\ref{eq:ADMM-nu})--(\ref{eq:ADMM-end}) are iteratively carried out until it converges.

\section{Analysis in the trash: Usage of whole $\tau$-regime data of $C(t,\tau/a)$ at a finite flow-time }\label{sec:All-tau-analysis}
It may be worth to see the sparse modeling analysis using all $\tau_i$ data at a finite flow-time.
The numerical codes written in C++, Fortran, and Julia are uploaded on the arXiv page of this paper.
The data of the smeared correlation function at $t/a^2=1.50$ is also included in the package.

\begin{figure}[h]
\begin{center}
\includegraphics[scale=0.55]{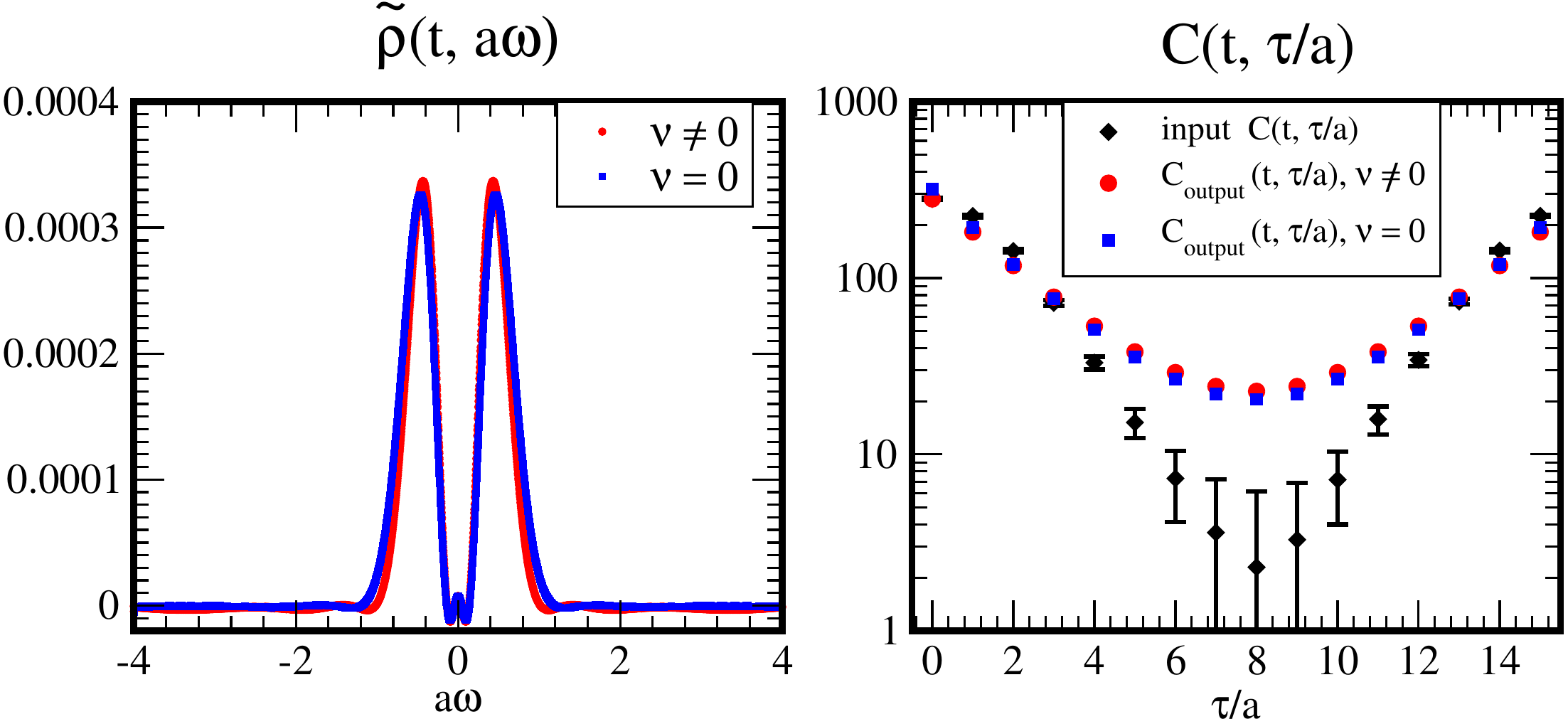}
\caption{(Left) the obtained spectral function (Right) the comparison plot between the input $C(t,\tau/a)$ and $C_{output}(t,\tau/a)$.}
\label{fig:trash-ana}
\end{center}
\end{figure}
Figure~\ref{fig:trash-ana} depicts the result of the spectral function in the left panel and the comparison plot between the input $C(t,\tau/a)$ and $C_{output}(t,\tau/a)$ in the right panel.
Here, the label $\nu \ne 0$ and $\nu =0$ denotes the results of the sparse modeling analyses with and without the sum rule, respectively.
In both results,  $\tilde{\rho}(t,a\omega)$ take the negative values near $\omega \approx 0$ even though we utilize the non-negativity as a constraint.
It numerically indicates that the analyses do not converge within a finite number of iterations in the ADMM routine  (here, we take $\mathcal{O}(10^9)$ iterations).

Furthermore, $C_{output}(t,\tau/a)$, which are reconstructed by the obtained spectral functions, far from the input one.
In particular, we can see that the slope in the log-scale is different between the input and the outputs.
It suggests that the input smeared correlation function can not be described by the kernel given by the hyperbolic cosine function in Eq.~(\ref{eq:kernel-SVD}).

\section{$N_\tau$ dependence of the singular values}\label{sec:App-sl}
The left panel in Fig.~\ref{fig:log-sl-tau-deps} show that the number of $s_l$ above $10^{-16}$ is the same with the number of the independent site in $\tau$ direction.
It tells us the number of the data-point on $N_\tau=16$ lattice is very sparse to resolve the information of the kernel  in double precision on a computer.
Here, we investigate how large lattice size (or how fine lattice spacing) are needed to analyze the kernel at this temperature with minimal loss of information.

For simplicity, here we consider the standard kernel (Eq.(\ref{eq:kernel-SVD})) instead of the reduced one,
\beq
K(\tau'_i,\omega'_j)  \equiv \frac{\cosh [\frac{\omega_j ' \Lambda }{2} (2 \tau_i' -1)]}{\cosh (\omega_j ' \Lambda /2)} \sqrt{\Delta \omega'}.
\eeq
Here, $\Lambda =  N_\tau a \omega_{cut} =\omega_{cut}/T $.
We assume that $\omega_{cut}$ in the physical unit is universal at a fixed temperature, so that $\Lambda$ is constant.
To increase the number of $s_l$ above $10^{-16}$, we have to take the larger $N_\tau$ and the finer $a$ at the fixed temperature.
Thus, it corresponds to taking the continuum extrapolation.
In the present paper, we have shown that  $a\omega_{cut}=4.0$ on $N_\tau=16$ lattice extent is a good choice to analyze the correlation function at $T=1.65T_c$, then we set $\Lambda=64$.

\begin{figure}[h]
\begin{center}
\includegraphics[scale=0.55]{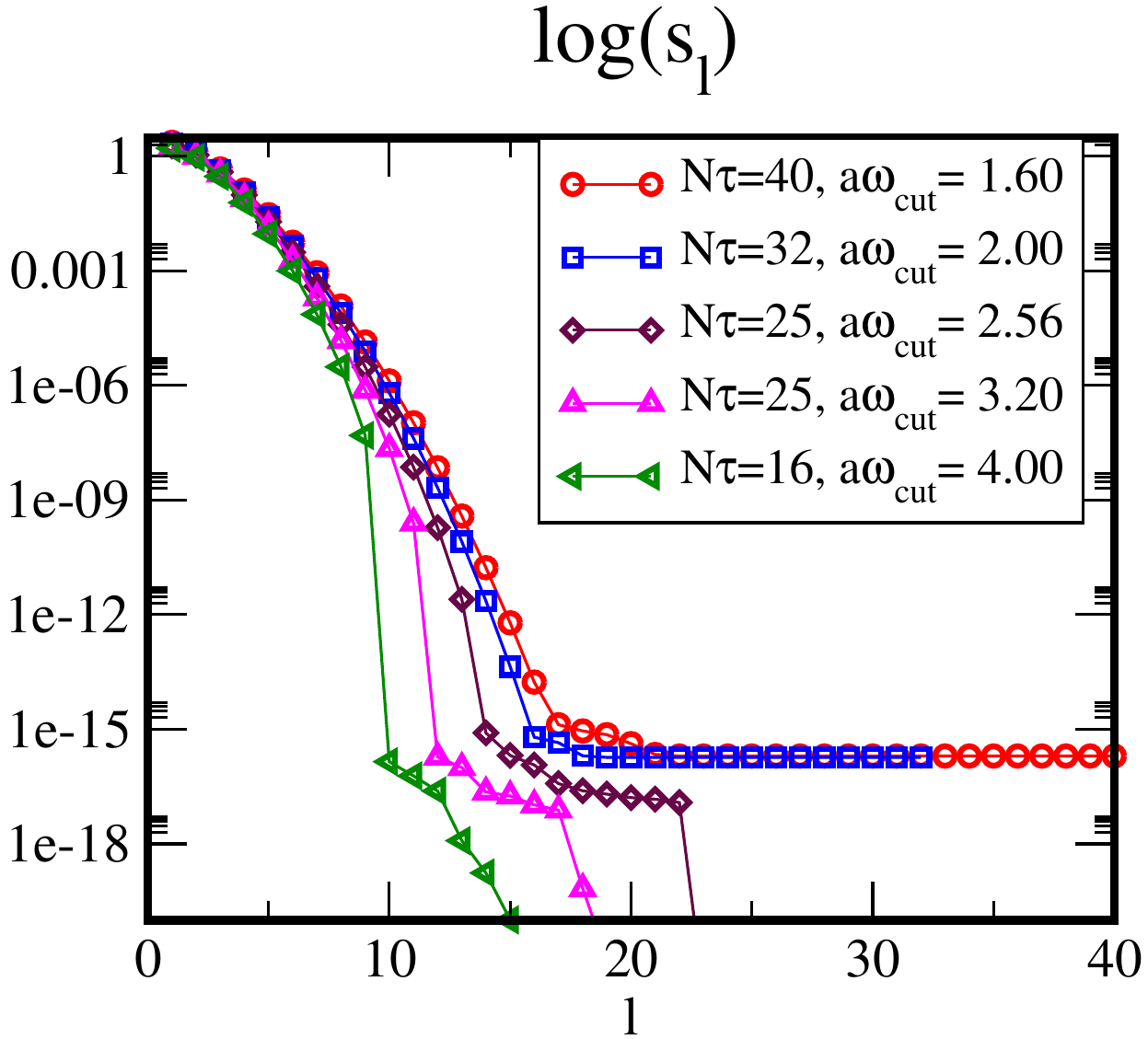}
\caption{$N_\tau$ dependence of the singular values for the kernel matrix. Here, we fix $N_\tau a \omega_{cut}=64.0$  }
\label{fig:log-sl-Ntau-deps}
\end{center}
\end{figure}
Figure~\ref{fig:log-sl-Ntau-deps} depicts the $N_\tau$ dependence of the singular values for the kernel matrix.
We see that the number of $s_l$ larger than $10^{-16}$ is almost saturated if we take $N_\tau \ge 32$.
Thus, it suggests that  the correlation function with the kernel above can be well described by $N_\tau \approx 32$ with minimal loss of information within the double precision.
In other words, the information will not increase even though we carry out the simulation on $N_\tau \gg 32$ in the double precision.

The minimal size of $N_\tau$ with minimal loss of the information depends on the temperature in the physical unit.
We expect that the lower temperature analysis needs the larger $\Lambda= \omega_{cut}/T$ and then the slope of $s_l$ becomes gentle~\cite{Chikano2018a,Chikano2018b,Li2019}.
Then, in the lower temperature, we need the larger lattice size to resolve the information of the kernel in the same precision.

\end{document}